\documentclass[aps,pra, twocolumn, preprintnumbers, groupedaddress, superscriptaddress,nofootinbib,10pt]{revtex4-1}
\usepackage[dvipsnames]{xcolor}
\usepackage[utf8]{inputenc}
\usepackage{url}
\usepackage{amsmath}
\usepackage{tikz}
\usepackage{amsfonts}
\usepackage{amssymb} 
\usepackage{bm}
\usepackage{graphicx}
\usepackage{array}
\usepackage{lipsum}
\usepackage{float}
\usepackage{multirow}
\usepackage{hhline}
\usepackage{tabularx}
\usepackage{subfigure}
\usepackage{graphicx}
\usepackage{afterpage}
\usepackage{epstopdf}
\usepackage{MnSymbol}
\usepackage[utf8]{inputenc}
\usepackage[hidelinks]{hyperref}

\usepackage{amsthm}

\newcommand{\cO}{\mathcal{O}}
\def\cV{\mathcal{V}}
\def\ct{\mathcal{T}}
\def\KB{\overline{K}_B}

\makeatother
\hypersetup{
    pdftitle={Constraints on Standard Model Constructions in F-theory},
    pdfauthor={Mirjam Cvetic, James Halverson, Ling Lin, Cody Long},
    pdfsubject={String Landscape, Standard Model, F-theory, Dark Sectors}
}

\begin{document}

\preprint{\texttt{CERN-TH-2020-049, UPR-1304-T}}

\title{
	Constraints on Standard Model Constructions in F-theory
	}

\author{Mirjam Cveti\v{c}} \affiliation{Department of Physics and Astronomy, University of Pennsylvania, Philadelphia, PA 19104-6396, USA}
  \affiliation{Center for Applied Mathematics and Theoretical Physics, University of Maribor, Maribor, Slovenia}
\author{James Halverson} \affiliation{Department of Physics, Northeastern University, Boston, MA 02115-5000 USA} 
\author{Ling Lin} \affiliation{Department of Theoretical Physics, CERN, 1211 Geneva 23, Switzerland}
\author{Cody Long} \affiliation{Department of Physics, Northeastern University, Boston, MA 02115-5000 USA} \affiliation{Department of Physics, Cornell University, Ithaca, NY 14853, USA}

\date{\today}

\begin{abstract}
\noindent We argue that the following three statements cannot all be true: 
(i) our vacuum is a type IIB / F-theory vacuum at moderate-to-large $h^{1,1}$,
(ii) the $\alpha'$-expansion is controlled via the supergravity approximation, \`{a} la the KKLT and LVS scenarios, and
(iii) there are no additional gauged sectors from seven-branes.
Since nearly all known globally consistent F-theory models with the exact chiral
spectrum of the Standard Model and gauge coupling
unification occur at
moderate $h^{1,1}$, this finding calls for new moduli stabilization scenarios or/and a rich seven-brane dark sector.
\end{abstract}

\maketitle

\section{Introduction} 
Explicitly realizing the Standard Model of particle physics, with all moduli stabilized and all experimentally measured couplings reproduced, is an important open problem in string theory that deserves immense effort. While some aspects related to moduli stabilization and supersymmetry
breaking are more difficult to achieve, others such as the gauge group and the chiral spectrum of the Minimal Supersymmetric Standard Model (MSSM) have been shown to have string embeddings \cite{Cvetic:2001nr,*Cvetic:2001tj,Braun:2005ux, *Bouchard:2005ag,*Braun:2005nv,*Bouchard:2006dn, *Anderson:2007nc, *Anderson:2009mh,  Cvetic:2015txa,Lin:2014qga, *Lin:2016vus, *Cvetic:2018ryq}, with the largest-to-date ensembles of global models with the exact chiral spectrum of the Standard Model being in F-theory \cite{Cvetic:2019gnh}.
For brevity, we henceforth refer to models in
the latter ensemble as MSSMs, despite the fact that the computation of Higgs pairs remains an open technical challenge.

Most four-dimensional F-theory compactifications exhibit gauge sectors much larger than the Standard Model, due to the generic presence of non-Higgsable clusters (NHCs) \cite{Grassi:2014zxa,*Morrison:2014lca}, which are gauge groups that do no admit a geometric Higgsing via a complex structure deformation.
Such NHCs occur generically \cite{Halverson:2015jua, Taylor:2015ppa, *Taylor:2015xtz, *Taylor:2017yqr, *Halverson:2017ffz, *Halverson:2017vde, *Wang:2020gmi} in F-theory bases with large $h^{1,1}$, which can be seen from the details of the topological transitions that move through moduli space from one elliptically fibered Calabi--Yau fourfold to another. As a vast majority of topological types of bases have large $h^{1,1}$, including the base believed to host the largest number of flux vacua~\cite{Taylor:2015xtz}, one concludes that a vast majority of F-theory compactifications have large gauge sectors, which should also apply to MSSM embeddings that likely exist at large $h^{1,1}$. 

However, there exists a class of F-theory bases for which there are no NHCs, namely the weak Fano bases. 
Such bases allow the construction of MSSMs, without necessarily requiring the presence of dark seven-brane gauge sectors. 
Toric weak Fano threefolds have $h^{1,1} \leq 35$, each corresponding to a fine, regular, star triangulation of a 3d reflexive polytope. 
Each of these geometries (provided a triangulation-independent numerical condition is satisfied) supports an MSSM construction without dark gauge sectors, by choosing all seven-brane stacks to wrap a particular homology class in the base, given by the anti-canonical divisors, and thus also ensuring  gauge coupling unification. 
This gives an estimated $\mathcal{O}(10^{15})$ MSSM compactifications in F-theory without dark seven-brane gauge sectors, the vast majority of which arising from a polytope with $h^{1,1}=35$, which is the largest value
amongst the weak Fano toric bases, but is moderate with respect to the ensembles of \cite{Halverson:2015jua, Taylor:2015ppa, *Taylor:2015xtz, *Taylor:2017yqr, *Halverson:2017ffz, *Halverson:2017vde}. Nevertheless,
our main conclusions already apply to these cases, and even more so to
cases with $h^{1,1}\gg 35$.

\medskip

Our main result is a non-trivial correlation between gauge couplings, dark sectors,
and control of the effective theory. In string theory language, this is a non-trivial
correlation between cycle volumes and tadpole cancellation in the supergravity approximation. 
Since we find the result physically intuitive, we briefly state it here.

Non-abelian gauge couplings arising from gauge sectors on seven-branes are inversely proportional to the volume, $\tau$, of the cycle that a given seven-brane wraps: 
\begin{equation}\label{eqn:gvo1l}
\frac{1}{\alpha}=\frac{4\pi}{g^2} =  \tau \, .
\end{equation}
Here the gauge coupling $g$ is fixed by $\tau$ at a fixed energy scale below the cutoff of the effective field theory, usually assumed to be the Kaluza--Klein scale of the compactification.
For example, in a GUT we have $\alpha_{3,2,1}\simeq \alpha_\text{GUT} \simeq 1/25$, corresponding to $\tau_{3,2,1}\simeq 25$.

For computational control of the effective theory, it is common to require that all cycles to have volume greater than some fixed cutoff, typically taken to be $\mathcal{O}(1)$. This assumption underlies both the KKLT \cite{Kachru:2003aw} and LVS \cite{Balasubramanian:2005zx} moduli stabilization scenarios, and the associated region of moduli
space is known as the stretched K\"ahler cone \cite{Demirtas:2018akl}. 
There, it was shown that at large $h^{1,1}$, in the regime of control at least one toric divisor tends to be large, often $\tau \gtrsim \cO(10^3)$, leading to an ultralight axion-like particle that can have significant cosmological implications \cite{Halverson:2019cmy}.

In this work we show that the large divisor also has significant implications for gauge sectors. 
Specifically, gauge sectors whose homology class has a contribution from this divisor necessarily have associated gauge coupling $\alpha \lesssim 1/1000$ within the stretched K\"ahler cone. 
If the visible sector (e.g., an MSSM or GUT) lives on seven-branes wrapped on such a cycle, then the UV gauge coupling is inconsistent with the observed values. 
In particular, wrapping the anti-canonical divisor, which receives volume contributions from \emph{all} toric divisors, leads to tensions if one relies on conventional moduli stabilization scenarios.
Concretely, by numerically minimizing the volume of the anti-canonical divisor in the stretched K\"ahler cone (a quadratic programming problem) at moderate $h^{1,1}$, we find that only ${\cal O}(10^4)$ of the models in \cite{Cvetic:2019gnh} have a realistic UV gauge coupling in the most common regime of control.

Given this correlation, it is natural to explore ways around it.
One is to wrap the MSSM or GUT seven-branes on small cycles (in homology), which correlate with 
smaller volume cycles. 
We show that, due to seven-brane tadpole cancellation, such constructions require the presence of additional seven-branes, often providing dark seven-brane gauge sectors\footnote{Dark sectors can also arise from D3-branes,
which are generally present in F-theory compactifications and can be uncharged under the Standard Model gauge symmetry. We focus entirely on dark sectors that arise from seven-branes.} and a rich cosmology. 
Alternatively, one could keep the visible sector on anti-canonical divisors but abandon the large $h^{1,1}$ assumption. However, this takes
one away from the bulk of the MSSMs in \cite{Cvetic:2019gnh}, which was dominated by geometries with $h^{1,1} = 35$. Finally, one could instead give up the restriction to remain in the stretched K\"ahler cone, requiring new moduli stabilization schemes that exhibit much more control
over non-perturbative effects.

\medskip

We therefore posit that generally the following three statements cannot all be true: i) our vacuum is a type IIB / F-theory vacuum at moderate-to-large $h^{1,1}$, ii) the $\alpha'$-expansion is controlled via the supergravity approximation, \`a la the KKLT and LVS scenarios, and iii) there are no gauged dark sectors from seven-branes. 
While in principle there can be exceptions, we will provide strong evidence that violation of this rule is rare, if it occurs at all.

This note is organized as follows.
In section \ref{sec:eft}, we revisit the key aspects of the effective field theory description and non-perturbative corrections of F-theory/IIB compactifications.
In section \ref{sec:dark_sectors}, we elaborate on the correlation between tuned visible sectors and forced dark sectors from seven-branes at large $h^{1,1}$.
For constructions where we can entirely avoid such dark sectors, we numerically analyze, in section \ref{sec:quad}, the bounds on the UV gauge couplings within the toric weak Fano landscape.
We close with some comments on challenges and benefits from abandoning one of the conditions in section \ref{sec:discussion}.

\section{Effective field theory considerations}\label{sec:eft}

We will consider ${\cal N}=1$ gauge theories that arise on seven-branes in F-theory compactifications, and will use the convention of unit string length.
Such a compactification is specified by an elliptically fibered Calabi--Yau fourfold $X \rightarrow B$, where the base $B$ are the physical extra dimensions of the dual type IIB spacetime (see \cite{Weigand:2018rez,*Cvetic:2018bni} for recent reviews).
We will study the requirement that the UV couplings of the Standard Model gauge groups take realistic values in the UV.
This is clearly model-dependent,
but we will take as a target $\alpha_{3,2,1}= \alpha_\text{GUT} = 1/25$ and will often find significant deviations from
these values for compactification within the stretched K\"ahler cone.

For control of the effective field theory we demand that all cycles in the K\"ahler threefold $B$ to be somewhat larger than appropriate powers of the string length. 
In this regime the gauge coupling $g_{\mathrm{UV}}$ is related to the volume $\tau_D$ of the seven-brane wrapping a divisor $D$~\cite{Grimm:2010ks}, via
\begin{equation}\label{eqn:gvol}
\frac{4\pi}{g_{\mathrm{UV}}^2} =  \tau_D \, ,
\end{equation}
where $\tau_D$ can be computed by the calibration
\begin{equation}
\tau_D = \frac{1}{2} \int\limits_D J \wedge J\, ,
\end{equation}
provided by the K\"ahler form $J$ on $B$. 

In an effective field theory arising in a string compactification on $B$, the tree-level action is corrected by Euclidean strings and branes wrapping appropriate-dimensional cycles in $B$. In an $\mathcal{N} = 1$ theory such corrections can be organized as to whether they correct the K\"ahler potential, superpotential, or higher-order F-terms. For instance, the tree-level K\"ahler potential for the K\"ahler moduli takes the form
\begin{equation}\label{eqn:ktree}
K = -2\, \mathrm{log}\cV\, ,
\end{equation}
where $\cV$ is the volume of $B$ computed as
\begin{equation}
\cV = \frac{1}{6} \int\limits_{B} J \wedge J \wedge J\, .
\end{equation}
Since the K\"ahler potential in $\mathcal{N} = 1$ theories does not appreciate any non-renormalization properties, it can receive a plethora of corrections. 
In particular, in the weakly-coupled orientifold case, worldsheet instantons wrapping a holomorphic curve $C$ are known to correct the $\mathcal{N} = 2$ prepotential, and are expected to descend to the $\mathcal{N} = 1$ IIB theory, providing corrections of the form~\cite{Hosono:1994av}:
\begin{equation}\label{eqn:nonp}
\Delta K \sim \frac{1}{\cV}\sum\limits_{n = 1}^{\infty} c_n e^{-2\pi n \sqrt{g_s} \mathrm{vol}(C)}\, ,
\vspace{.2cm}
\end{equation}
where $g_s$ is the string coupling, taken to be a constant in the weakly-coupled type IIB orientifold limit of F-theory. 
The constants $c_n$ are (related to) the Gromov--Witten invariants, that in essence count unique representatives of the curves in the class $[n C]$. 
 
In order to treat Eq.~\ref{eqn:nonp} as a small correction to Eq.~\ref{eqn:ktree}, we need to enforce that $\mathrm{vol}(C) \gtrsim c$, for some constant $c$, which depends on the functional form of $\cV$, the constants $c_n$, and the allowed error in the computation of relevant physics.
Restricting all curves volume to be greater than some threshold $c$ is known as \textit{restricting to the $c-$stretched K\"ahler cone}, as defined in~\cite{Demirtas:2018akl}. 
Typically $c$ is taken to be $\mathcal{O}(1)$. However, restricting to this cone is in general not a sufficient condition to be in a regime of control; for instance, the leading Gromov--Witten invariants can be quite large ($\mathcal{O}(10^3)$ in the simplest example~\cite{Candelas:1990rm}). In addition, Euclidean D3-branes (ED3s) wrapping 4-cycles $\Sigma$ can provide corrections to both the K\"ahler and superpotential, and the corrections to the Lagrangian $\mathcal{L}$ are schematically
\begin{equation}
\Delta \mathcal{L} \sim e^{-2\pi \mathrm{vol}(\Sigma)}\, .
\end{equation}
ED3 corrections to the superpotential are well-studied and are important for moduli stabilization. On the other hand, ED3 corrections to the K\"ahler potential are relatively un-studied, as the ED3 wraps a volume-minimizing surface (in its class), whose volume is difficult to compute, but can perhaps be bounded by quantum gravity considerations~\cite{Demirtas:2019lfi}. 
In any case, we expect uncontrolled corrections if any surface volume $\tau_i$ satisfy $\tau_i \lesssim 1$. 
It is important to note that it may be possible to shrink some curves and surfaces down to small volume, without spoiling the validity of the effective description, such as certain orbifold limits. We will therefore use the stretched K\"ahler cone, and the analogous restriction on divisor volumes, as a proxy for control, and study the consequences.

The central observation of~\cite{Demirtas:2018akl} is that, as $h^{1,1}(B)$ grows large, these restrictions force some four-cycle volumes $\tau_i$ to be very large, due to the fact that the K\"ahler cone becomes very narrow at large $h^{1,1}$. 
In the case of type IIB compactified on a Calabi--Yau orientifold $B$ this implies the presence of an essentially massless axion. However, from Eq.~\ref{eqn:gvol} one can see that this growth will affect the distribution of gauge couplings at large $h^{1,1}$: large cycles will force small UV gauge couplings for any gauge groups supported on seven-branes wrapping the large cycle. Clearly this can provide an obstruction to realizing the Standard Model, with the UV correct gauge coupling, on seven-branes wrapping such cycles.

Our focus in the remainder of this manuscript will be on understanding the relationship between the stretched K\"ahler cone, $h^{1,1}$, and the appearance of seven-brane dark sectors. 
Our findings suggest that, by tuning a visible sector on seven-brane stacks, one may choose two, but not three, of the following simultaneously:
\begin{enumerate}
\item Control of the theory via the stretched K\"ahler cone.
\item The absence of seven-brane dark sectors.
\item Large $h^{1,1}$.
\end{enumerate}

The first item is necessary for control of the low energy effective theory given current understanding, but not a necessary condition for the theory's existence: there is no fundamental principle stating that the string vacuum corresponding to our universe is weakly-coupled (in $g_s$ or $\alpha^\prime$).
The second is one of phenomenological interest, for as the number of additional sectors grows, the number of possibilities for model building grows; however, at the same time the possible constraints from experiments become more stringent (cf.~\cite{Halverson:2019cmy, Halverson:2016nfq, Halverson:2019kna}). Finally, there is an overwhelming amount of evidence that majority of both geometries and vacua exist at moderate-to-large $h^{1,1}$, from a flat counting as well as a cosmological perspective~\cite{Taylor:2015ppa, *Taylor:2015xtz, *Halverson:2017ffz, *Taylor:2017yqr, Carifio:2017nyb}.
That is, in the absence of an extremely selective yet-unknown measure factor, one should expect that our string vacuum is at moderate-to-large $h^{1,1}$.

\section{Ubiquity of dark sectors}\label{sec:dark_sectors}

To correctly produce the UV gauge coupling one might attempt to wrap seven-branes carrying the visible sector only on small divisors, whose volumes define a boundary of the stretched K\"ahler cone.
These divisors will have the smallest possible volume while maintaining perturbative control, and can in principle (depending on the topology of $B$) support the visible sector with realistic gauge couplings.
However, with increasing $h^{1,1}(B) \equiv h^{1,1}$, the presence of such gauge sectors on small divisors comes at the expense of ``dark'' sectors, which may not really be dark, as we will explain now.

The origin of these additional gauge sectors lies in the cancellation of seven-brane tadpoles.
Namely, the divisors $W_i$ wrapped by seven-branes must satisfy
\begin{align}
	\sum_i \delta_i W_i = [\Delta] = 12 \KB \, ,
\end{align}
where $\delta_i$ is the vanishing order of the discriminant $\Delta$ of the elliptic fibration over $W_i$, and $\KB$ is the anti-canonical class of $B$.
Certain choices of irreducible divisors $D_\text{vis}$ to support the visible gauge sector (be it a GUT or directly the Standard Model) can force the presence of additional gauge sectors on ``nearby'' rigid divisors.
``Nearby'' in this context means that these divisors will intersect $D_\text{vis}$, which can lead to matter charged both under the visible and the additional gauge symmetries.\footnote{Of course, the precise spectrum in F-theory compactifications to 4d will depend on the flux background.
However, a detailed quantitative study of them will require case by case analyses of each individual geometry, which is beyond the scope of this work.
}
Their presence is a consequence of the required geometric tuning to obtain the visible gauge sector in the first place: Higgsing along these bi-charged matter corresponds to the inverse geometric deformation, which must break both the visible and the additional sector. 

Due to these matter states, these additional sectors could have non-trivial interactions with
the visible sector. The matter serves as messengers to a gauged dark sector
and may be constrained by LHC searches for vector-like quarks and leptons. 
Even if those matter states can be lifted by suitable fluxes or scalar field VEVs, there could be additional constraints arising from the overproduction of
dark glueballs \cite{Halverson:2016nfq}.
Of course, the details of these depends on the geometry of the base $B$.
For example, on $B=\mathbb{P}^3$, there are no rigid divisors, and consequently, any tuned visible sector will not lead to additional gauge sectors.
However, even in the regime of low $h^{1,1}$, mathematical and phenomenological consistency conditions may require the existence of seven-brane dark sectors (see, e.g., \cite{Cicoli:2012vw, *Cicoli:2013mpa, *Cicoli:2013cha}), whose numbers will grow with with $h^{1,1}$.

To make these statements precise, we will focus on the class of weak Fano toric spaces $B$,
which is the context of the recently found large ensemble of three-family Standard Models \cite{Cvetic:2019gnh}.
These spaces can be characterized by a three dimensional reflexive lattice polytope $\Diamond \subset N \cong \mathbb{Z}^3$ and a (fine regular star) triangulation.
Each vertex (except the single internal one) $v_i$ corresponds to a toric divisor $D_i = \{x_i = 0\} \subset B$.
Any effective divisor is linearly equivalent to a \emph{positive} linear combination of toric divisors $D_i$, $i=1,..., h^{1,1} + 3$.

Given any effective divisor $D = \sum_i a_i D_i$, there is a simple combinatorial formula that computes a basis for $H^0(B, \mathcal{O}_B(D))$ in terms of monomials formed out of the toric coordinates $x_i$.
Let 
\begin{align}\label{eq:dual_polytope_of_D}
	\Diamond_D^* = \{ m \in M\, | \, \forall i: \langle m, v_i \rangle \geq -a_i\} \, ,
\end{align}
where $M$ is the dual lattice of $N$.\footnote{Given a basis choice of $N \cong \mathbb{Z}^3$, one can simply take $M \cong \mathbb{Z}^3$, with $\langle \cdot, \cdot \rangle$ being the standard dot product.}
Then every $m \in \Diamond_D^*$ defines a monomial
\begin{align}\label{eq:monomials_combinatorial}
	d_m = \prod_i x_i^{\langle m, v_i \rangle + a_i} \, ,
\end{align}
such that any global section of $\mathcal{O}_B(D)$ is a polynomial
\begin{align}
	s = \sum_{m \in \Diamond_D^* } \lambda_m \, d_m \, \text{ with } \, \lambda_m \in \mathbb{C} \, .
\end{align}
Since $D = \{ s= 0\}$ for a suitable global section $s \in H^0(B, \mathcal{O}_B(D))$, it is clear that $D$ necessarily factorizes if all monomials $d_m$ share at least one common factor.

The case $D = \KB = \sum_i D_i$ deserves a detailed discussion.
In this case, $\Diamond_{\KB}^* \equiv \Diamond^*$ is just the dual polytope of $\Diamond$.
Because of the reflexivity requirement for $\Diamond$, each two-dimensional facet $F$ of $\Diamond$ has an associated point $m_F \subset M$ that is a vertex of $\Diamond^*$, satisfying $\langle m_F, v \rangle = -1$ for all $v \in F$.
Moreover, reflexivity implies that $\Diamond^*$ contains an interior point $m_0$ with $\langle m_0, v \rangle = 0$ for all $v \in \Diamond$.
As a consequence, a generic representative of $\KB$ is irreducible, because the monomial $d_{m_0} = \prod_i x_i$ contains a factor of every toric coordinate, but --- since any $v_i \in \Diamond$ is contained in at least one facet $F$ --- there is also a monomial $d_{m_F}$ without any $x_i$ factor, since $\langle m_F , v_i \rangle + a_i = -1 + 1 = 0$.
Incidentally, the irreducibility of $\KB$ guarantees that there are no non-Higgsable clusters in F-theory compactifications on bases $B$ of this type.

Let us briefly summarize the results that follow. We will generally be interested in tuning gauge groups on divisors in $B$, and checking how $\Delta^\prime$ further factorizes as a result of this tuning. We will first consider tuning a Standard Model gauge group along a prime toric divisor. The location of the prime toric divisor, that is whether it is a vertex, internal to an edge, or internal to a face, will analytically determine a minimal factorization of $\Delta^\prime$ into at least $I_1$ loci, which in particular interesting examples are enhanced to non-abelian gauge sectors. In particular, tuning a gauge group on divisor corresponding to a vertex, bounding edges $e_i$ and faces $F_j$, forces $I_1$ loci on all toric divisors corresponding to points in the strict interior of $e_i$ and $F_j$. In a similar vein, tuning a gauge group on a divisor corresponding to a point interior to an edge, bounding faces $F_j$, forces $I_1$ loci on all other divisors corresponding to points interior to that edge, as well as divisors corresponding to points on the strict interior to $F_j$. Finally, tuning a gauge group along a divisor corresponding to a point interior to a face forces an $I_1$ locus on all divisors corresponding to points interior to that face.

\subsection{Dark sectors from the Standard Model on prime toric divisors}

We now consider $I_1$ loci forced to occur on divisors corresponding to points interior to faces. 
We emphasize that the forced fibers are often more singular than $I_1$, but studying forced
$I_1$ fibers is a convenient way to proceed with the analysis.

Suppose there is a facet $F \subset \Diamond$ containing $v_k$ and another \emph{interior} point $v_\text{int} \neq v_k$.
Note that the toric divisor associated to $v_\text{int}$ is rigid on $B$.
As facet internal points, the only point in $\Diamond^*$ with $\langle m, v_\text{int} \rangle = -1 $ is $m_F$, the point dual to the facet $F$; it also satisfies $\langle m_F, v_k \rangle = -1$.
Now consider the divisor $D = \KB - D_k$, whose monomials are determined by points $m \in \Diamond^*_D$, i.e., must have $\langle m, v_k \rangle \geq 0$.
Since $m_F$ is clearly not in $\Diamond^*_D$, this means that we must have $\langle m, v_\text{int} \rangle \geq 0$ for all $m \in \Diamond_D^*$.
Hence, the exponent of $x_\text{int}$ in every monomial $d_m$ is $\langle m, v_\text{int} \rangle + 1 \geq 1$.
In other words, any representative in the divisor class of $D = \KB - D_k$ is reducible and contains components $\{x_\text{int}=0\}$ for any other point internal to the same facet $F$ containing $v_k$.  

A similar argument applies to $v_k$ on an edge, and $v_\text{int}$ in the strict interior of the same edge. 
Tuning a gauge group on the prime toric divisor corresponding to $v_k$ requires an $I_1$ locus on the divisor corresponding to $v_\text{int}$, which can be easily seen from linearity: 
the presence of a gauge group on $D_k$ requires $\langle v_k , m \rangle \geq 0 \, $ for all $ m \in \Diamond^*_D$ with $D = \KB - D_k$. 
If we assume no gauge group on the divisor corresponding to $v_\text{int}$ this implies that for at least one $m \in \Diamond^*_D$ we have $\langle v_\text{int}, m \rangle = -1$, but by linearity we must have $\langle v_l , m \rangle \leq -2$ for some $v_l$ on the edge, on the opposite side of $v_\text{int}$ from $ v_k$, which is a contradiction.
Therefore tuning a gauge group on $v_k$ on an edge requires at least an $I_1$ locus on all divisors corresponding to interior points to that edge.

To apply these facts to F-theory models, suppose we engineer a visible sector (a GUT or non-abelian part of the Standard Model group) on a \emph{toric} divisor $D_\text{vis}$ wrapped by $\delta$ seven-branes.
These setups will have the smallest possible value for $\tau_{D_\text{vis}}$ within the stretched K\"ahler cone, as all effective divisors are generated by toric divisors.
Since we have $\overline{K}_B = \sum_i D_i$ for toric spaces, there must be additional seven-branes wrapping the residual discriminant with class
\begin{align}
	[\Delta'] \equiv [\Delta] - \delta D_\text{vis} =  12\,\sum_{i \neq \text{vis}} D_i + (12 - \delta) D_\text{vis} \, .
\end{align}
The corresponding polytope in the dual lattice is then
\begin{align}\label{eq:dual_polytope_residual_disc}
\begin{split}
	\Diamond^*_{\Delta'} = \{ m \in M \, | \, & \langle m, v_i \rangle \geq -12 \text{ for } i \neq \text{vis} \, , \\
	& \langle m ,v_\text{vis} \rangle \geq \delta-12 \} \, .
\end{split}
\end{align}
This situation is now similar to the previous example.
First, for any \emph{facet internal} point $v_\text{int} \neq v_\text{vis}$ sharing the same facet $F \subset \Diamond$, there is precisely one\footnote{
Since $\Diamond$ is reflexive, $S := \{ m \in M\otimes_{\mathbb{Z}} \mathbb{R} \, |\, \forall v \in \Diamond : \langle m, v \rangle \geq -1  \}$ is the convex hull of $\Diamond^* \subset M$.
If there was another vertex $\tilde{m} \neq 12 m_F$ with $\langle \tilde{m}, v \rangle = -12$ for all $v\in F$, $m' \equiv \frac{1}{12} \tilde{m} \in M\otimes_{\mathbb{Z}} \mathbb{R}$ would satisfy $\langle m' ,v \rangle = -1$ for all $v \in F$, placing $m'$ on the \emph{boundary} of $S$.
By linearity, the line between $m_F$ and $m'$ must then also be on the boundary of $S$, and thus either be inside a facet, or be itself an edge of $S$, and hence also in a facet or an edge of $\Diamond^*$.
This facet or edge must end on at least one other (integral) vertex $\hat{m}$ in $\Diamond^*$ different than $m_F$, which by linearity also satisfies $\langle\hat{m}, v \rangle = -1$ for all $v \in F$.
This is a contradiction to the fact that $\Diamond^*$ (as dual to a reflexive polytope) has exactly one vertex saturating the inequality $\langle m, v \rangle \geq -1$ for all $v \in F$.
} vertex $m_{F,\Delta} = 12 \times m_F \in \Diamond^*_\Delta$ with $\langle m_{F,\Delta}, v_\text{vis} \rangle = \langle m_{F,\Delta}, v_\text{int} \rangle = -12$.
As this point is not in $\Diamond_{\Delta'}^*$ per Eq.~\ref{eq:dual_polytope_residual_disc}, it follows analogously that every monomial must have a positive power of $x_\text{int}$.

Note that for the existence of additional gauge sectors on a common factor $\{x_\text{int}=0\}$, the power of $x_\text{int}$ in all monomials of $\Delta'$ must be at least 2 or 3 (depending on additional details of the elliptic fibration that we will neglect here).
This now depends, in addition to the choice $(D_\text{vis}, \delta)$ of tuning, also on the details of the polytope $\Diamond$, e.g., the number and relative positions of the interior points $v_\text{int}$ in the same facet as $v_\text{vis}$.

For concreteness, let us consider the polytope $\Diamond_8$ (in the numeration of \cite{Kreuzer:1998vb}), which is one of two polytopes that dominated the ensemble of Standard Models constructed in \cite{Cvetic:2019gnh}.
This polytope has 38 points (which is the maximum amongst 4319 polytopes classified in \cite{Kreuzer:1998vb}), i.e., the corresponding bases have $h^{1,1}(B) = 35$.
$\Diamond_8$ has four facets, three of which are exchanged by a $\mathbb{Z}_3$ symmetry of the polytope; these three have two internal points, while the fourth facet has 10 internal points.
One can then check explicitly that for the minimal requirement of the visible sector to contain an $SU(3)$, i.e., a tuned $I_3$ fiber with $\delta = 3$ or type IV fiber with $\delta = 4$, one will always find at least two $I_2$ singularities, i.e., $SU(2)$ sectors on facet interiors.
For an $SU(5)$ GUT with $\delta = 5$, one finds at least two $SU(4)$ sectors or even more factors of lower rank.
Similar enhancements going beyond the desired gauge sector also occurs on bases defined by other polytopes.
In general, the bigger the facets of the polytopes are (i.e., the more rigid divisors $B$ has), the more additional gauge sectors appears. 
For example, there exists one other polytope with 38 points which has a similar number of enhancements in the presence of a tuned GUT or MSSM gauge group.

\subsection{Dark sectors from visible sector on square-free divisors}
To attempt to avoid such enhancements while still maintaining perturbative control within the stretched K\"ahler cone, we can also consider the case of divisors ``larger'' than prime torics, where $D_\text{vis}$ is a \emph{square-free} divisor, i.e., a divisor where all $a_i$ are either 0 or 1.
Clearly, these will be the smallest divisors after the toric ones, which are themselves a special case of square-free divisors.
However, such divisors will in general not be irreducible by themselves, and thus lead to additional gauge factors in the effective theory.
Necessary conditions for non-factorization for square-free divisor $D$ can be derived simply be inductively including the toric components of $D$. 
Let $\ct_D$ be the simplicial complex corresponding to $D$, induced by the triangulation of $\Diamond$. One finds~\cite{Braun:2017nhi}:
\begin{equation}\label{eqn:genera}
h^0 (B, \mathcal{O}_B (D))  = 1 + \sum\limits_v g(v)  + \sum\limits_e g(e)  + \sum\limits_f g(f) \, , 
\end{equation}
where $v,e,f$ are complete vertices, edges, and faces of $ \Diamond$ included in $\ct_D$, and $g$ is the generalized genus of an $n$-face $\Theta$, computed by the number of strict interior points $l^{*}(\Theta^\star)$ of its dual face in the dual polytope $ \Diamond^\star$: $g(\Theta) = l^{*}(\Theta^\star)$ (for a face $f$ we define $g(f) = 1$, since it is a facet of $ \Diamond$, dual to a vertex of $ \Diamond^\star$).\footnote{In the case that $D$ is an anti-canonical divisor there is an additional additive factor of 1 in this formula from the fact that the entire boundary of the 3-polytope is included.}  This immediately implies that if $\ct_D$ includes any points interior to an edge or a face, it must include all points on that edge or face in order not to factorize. In addition, the genus of the edge or face must be non-zero to avoid factorization. 

As a concrete example, consider $\Diamond_8$, whose vertices are given by 
\begin{equation}
[[-1,-1,-1],[-1,5,-1],[-1,-1,5],[1,-1,-1]]\,. 
\end{equation}
This polytope is a simplex. $ \Diamond_8$ exhibits a $\mathbb{Z}_3$ symmetry, and has a single large face with 28 points on it, and 5 points interior to each of its edges. The three edges emanating off of the large face have one interior point each, and the remaining three faces have two interior points per each one. Let us now consider a general square-free divisor $D$ on a toric variety $B$, corresponding to a simplicial decomposition $\mathcal{T}_D$ of $\Diamond_8$. To compute the number of global sections of $\mathcal{O}(D)$ we will use Eq.~\ref{eqn:genera}. The structure of $\Diamond_8$ makes it particularly simple to count the number of square-free divisors with deformation. In $ \Diamond_8$, the only vertex with a non-zero genus is $[1,-1,-1]$. All edges are genus-zero, and all faces are genus one. Recall that for a square-free divisor, one must include all points on a face or edge for the genus of that face or edge to contribute to the number of global sections. Note that all faces and all edges have at least one interior point. 
Any simplicial complex $\ct_D$ will correspond to a reducible divisor $D$ if $\mathcal{T}_D$ includes any point on a face (not necessarily interior to), without including that face, unless that point is $[1,-1,-1]$, or $\ct_D$ is a prime toric divisor. First, there are 38 square-free toric divisors, which are irreducible and do not factorize. If our complex $\ct_D$, corresponding to our square-free divisor $D$, includes the point $[1,-1,-1]$ but none of the facets it bounds, then the only way to increase $H^0 (B,\mathcal{O}_B(D))$ is to include the face $F$ opposite $[1,-1,-1]$, and one or more connecting edges to $F$ to make $\ct_D$ connected, but such a divisor necessarily factorizes since the edges are genus-zero.  Finally, one can consider the combinatorial ways of adding in full faces, of which there are $2^4$. A simple counting yields there are $38+ 2^4 = 54$ square-free divisors that satisfy this property. The total number of non-trivial square-free divisors on $B$ is $2^{38} -1 = 274877906943$, and so the irreducible square-free divisors on $B$ constitute a fraction of $2 \times 10^{-10}$ of all square-free divisors. Therefore most choices of square-free homology class of the standard model will produce additional standard-model-like dark sectors.

The prospect of additional forced $SU(3)\times SU(2)$ sectors is potentially interesting 
for the $N$-naturalness scenario \cite{Arkani-Hamed:2016rle}, though F-theory topology and its relation
to tadpole cancellation bounds the number of such sectors to not be too high.

\subsection{The \texorpdfstring{\boldmath{$\mathbb{P}_{F_{11}}$}}{F11} model and \texorpdfstring{\boldmath{$\KB$}}{KB} }

As a final concrete example, let us consider the $\mathbb{P}_{F_{11}}$ model \cite{Klevers:2014bqa} which realizes the exact Standard Model gauge group in the absence of any additional enhancement.
The (resolved) elliptically fibered Calabi--Yau geometry is described by the hypersurface
\begin{align}
P &= s_1 e_1^2 e_2^2 e_3 e_4^2 u^3 + s_2 e_1 e_2^2 e_3^2 e_4^2 u^2 v + s_3 e_2^2 e_3^3 u v^2 \nonumber \\
& +s_5 e_1^2 e_2 e_4^3 u^2 w + s_6 e_1 e_2 e_3 e_4 u v w + s_9 e_1 v w^2 = 0\, .
\end{align}
The coefficients are sections of the following bundles:
\begin{align}\label{eqn:secs}
\begin{array}{ll}
[s_1] = 3 \KB - S_7 - S_9\, , & [s_2] = 2\KB - S_9\, , \\ [1ex]
[s_3] = \KB + S_7 - S_9\, , & [s_5] = 2\KB - S_7\, , \\ [1ex]
[s_6] = \KB\, , & [s_9] = S_9 \, .
\end{array}
\end{align}
Here different choices of the classes $[s_i]$ correspond to topologically inequivalent MSSM fibrations over a fixed base. 
Every such fibration, labeled by a choice of divisor classes $(S_7, S_9)$, will have an $I_2$ singularity corresponding to an $SU(2)$ gauge symmetry over $\{s_3=0\}$, and an $I_3$ singularity, i.e., an $SU(3)$ gauge symmetry, over $\{s_9=0\}$.
Generally, $(S_7,S_9)$ must be chosen such that the classes $[s_i]$ are effective on $B$.
As we have seen above, however, requiring that the visible sector does not factorize will limit the choices of $(S_7,S_9)$.
Moreover, even if $s_3$ and $s_9$ do not factorize, the residual discriminant $\Delta'$ (which itself is a complicated polynomial in the $s_i$) can factorize through the factorizations of the other coefficients.
For example, if $s_1 = s \, s_1'$ and $s_5 = s \, s_5'$, then $\Delta' \sim s^2$, signaling an $I_2$ fiber over $\{s=0\}$.

In general, the precise number of additional sectors is highly-dependent on the base geometry. 
However, as utilized in~\cite{Cvetic:2019gnh}, the choice $S_{7,9} = \KB$ always provides an effective, irreducible solution to Eq.~\ref{eqn:secs}, as this choice renders all classes $[s_i]$ in Eq.~\ref{eqn:secs} to be the anti-canonical, which as we discussed above is always irreducible for a weak Fano toric threefold.
Despite this rather restrictive choice, there are around $\mathcal{O}(10^{15})$ compactifications based on the $\mathbb{P}_{F_{11}}$ fibration, which leads to a globally consistent three-family MSSM-like effective field theory with no additional seven-branes gauge sectors \cite{Cvetic:2019gnh}.

Given these considerations, it is natural to relate the gauge coupling(s) of the visible sector to the volume of the anti-canonical divisor $\KB$ of the base $B$ in F-theory compactifications.
Before we present a quantitative analysis for the ensemble of toric weak Fano bases, let us provide one further argument, independent of additional gauge enhancements and divisor (ir-)reducibility, why the volume of $\KB$ is interesting for phenomenological discussions.

\subsection{Couplings of \texorpdfstring{\boldmath{$U(1)$}}{U(1)}s}

The anti-canonical divisor $\KB$ of the base naturally appears in the context of seven-brane $U(1)$ gauge symmetries in F-theory.
Geometrically, they are described by rational sections $\sigma$ of the elliptic fibration \cite{Morrison:1996pp}.
The couplings of each $U(1)_A$ factor is tied to the volume $\tau_{b_A}$ of the so-called height pairing divisor $b_A \subset B$ associated to the section $\sigma_A$ generating $U(1)_A$ \cite{Park:2011ji,*Morrison:2012ei}.
As explained in \cite{Cox1979,*Lee:2018ihr,*Lee:2019skh}, the divisor takes the form
\begin{align}
	b_A = \frac{1}{m_A^2} (2 \KB + \eta_A) \, ,
\end{align}
where $\eta_A$ is an effective divisor and $m_A \in \mathbb{N}$ is a positive integer.
Both quantities depend on details of the fibration structure; however, we can provide an upper bound for the coupling,\footnote{For non-abelian couplings (cf.~Eq.~\ref{eqn:gvol}) there is an additional factor of 2, stemming from different normalizations of Lie algebra generators in particle physics and in algebraic geometry.}
\begin{align}
	g_A^2 = \frac{2\pi}{\text{vol}(b_A)} = \frac{2\pi m_A^2}{\text{vol}(2 \KB + \eta_A)} \leq \frac{\pi m_A^2}{\text{vol}(\KB)} \, .
\end{align}

The value of $m_A$ depends on how $\sigma_A$ intersects codimension one fibers of the elliptic fibration \emph{differently} than zero section $\sigma_0$ \cite{Cox1979,*Lee:2018ihr,*Lee:2019skh}.
For fibrations with a single non-abelian gauge factor over the discriminant locus $W \subset \Delta$, the integer $m_A$ is determined by requiring that $m_A \times \sigma_A$ \emph{in the Mordell--Weil group law} intersects the same Kodaira fiber component as $\sigma_0$ (i.e., the affine node over $W$).
In the presence of multiple discriminant loci $W_i$ carrying Kodaira fibers, $m_A$ is the smallest positive integer such that $m_A \times \sigma_A$ intersects the affine node over all $W_i$.

The different intersection pattern between $\sigma_A$ and $\sigma_0$ is also intimately tied to the global gauge group structure \cite{Cvetic:2017epq}.
In particular, for fibrations realizing the Standard Model gauge group $[SU(3) \times SU(2) \times U(1)_A]/\mathbb{Z}_6$, the section $\sigma_A$ intersects (up to a symmetry for $SU(3)$) the unique non-affine nodes of $SU(2)$ and $SU(3)$.
In the absence of any other non-abelian gauge algebras, e.g., as in \cite{Cvetic:2019gnh}, we thus have $m_A = 6$, and therefore
\begin{align}
	g_A^2 \leq \frac{36\pi}{\text{vol}(\KB)}.
\end{align}
Thus, the hypercharge coupling in direct realizations of the MSSM would always be sensitive to the volume of $\KB$.

One might wonder if additional non-abelian gauge sectors, which we just argued are generically present, could lead to a larger $m_A$ and thus increase this bound.
While this is possible in principle, the list of gauge algebras with such a desired effect is limited.
To have $m_A > 6$, the fiber structure must be such that no $6 \times \sigma_A$ does not intersect the affine node of the codimension one fiber associated with the additional gauge factor $G'$ \cite{Cox1979,*Lee:2018ihr,*Lee:2019skh}.
Because of the connection between the intersection patterns of sections and the global gauge group, this translates into the condition that the order of the fundamental group of $G'$ is \emph{not} a divisor of $6$.
This only applies to $G' = SU(n)$ with $n>3$ (and $n \neq 6$) or $G' = SO(2k)$, $k \geq 6$.
In the latter case, the maximal ``enhancement'' in $m_A \rightarrow m_A'$ is a factor of 2 (since $\pi_1 ( SO(2k)) = \mathbb{Z}_4$ or $\mathbb{Z}_2 \times \mathbb{Z}_2$), which is not a significant increase.
In the concrete examples considered above, where additional gauge sectors arose from an MSSM tuned over prime torics on bases $B$ associated to the polytope $\Diamond_8$, $m_A$ does not change at all since the additional gauge sectors are only $SU(2)$ or $SU(3)$.

In summary, we have argued that in concrete F-theory compactifications realizing the Standard Model on 7-branes, without additional dark sectors, the gauge couplings are always related to the volume $\tau_{\KB}$ of the anti-canonical divisor of $B$. In addition, if the $U(1)$ is realized by a section of the elliptic fibration, its gauge coupling is always related to $\tau_{\KB}$, independent of the model.
In the following, we will see that this severely limits compactifications with realistic coupling values within the stretched K\"ahler cone.

\section{The Distribution of Pure MSSM Gauge Couplings}\label{sec:quad}

We now analyze gauge couplings in the ensemble of weak Fano toric bases, with the MSSM  or the GUT supported on $\KB$. 
Recall from Eq.~\ref{eqn:gvol} that the gauge coupling of a gauge group supported on a seven-brane wrapping a divisor $D$ is determined by its volume $\tau_D \equiv \tau$. In this construction, each gauge group is supported on an anti-canonical divisor, whose volume is the sum of the volumes of the toric divisors:
\begin{equation}
\tau = \sum\limits_i \tau_i\, .
\end{equation}
For realistic gauge couplings, we require $\tau \simeq 25$. In the stretched K\"ahler cone, we require that all curves $C$ have $\mathrm{vol}(C) \geq 1$, and in general, control of the theory also requires $\tau_i \gtrsim 1$, which for large numbers of moduli is in tension with the value $\tau \simeq 25$; even a single large divisor spoils the UV gauge coupling requirement. Our first task is then to check which geometries allow for $\tau \simeq 25$ in the stretched K\"ahler cone. Each geometry corresponds to a triangulation of one of 4319 3d reflexive polytopes. As the number of triangulations is expected to be $\mathcal{O}(10^{15})$, enumerating all triangulations is beyond the scope of this work, and so we will consider a single triangulation per polytope.

Given a geometry, our task is then to compute the maximum gauge coupling achieved within the stretched K\"ahler cone. This is performed by minimizing the volume of the anti-canonical divisor, within the stretched K\"ahler cone. We must therefore solve
\begin{equation*}
\{ \mathrm{minimize}\, \, \mathrm{vol}(\KB) \, \, | \, \, \text{vol}(C) \geq 1 \, \, \forall \, \text{effective } \, C \subset B \}\, .
\end{equation*}
We perform this minimization using \textbf{Mathematica}. The minimization
problem itself is technically a non-convex global quadratic program, which in its general form is NP-hard.
This provides a new instance of computational complexity arising in string theory,
in addition to those of \cite{Denef:2006ad, *Cvetic:2010ky, *Halverson:2018cio, *Halverson:2019vmd}. 
Due to NP-hardness, (in the absence of additional structure) at moderate-to-large $h^{1,1}$, it is difficult to check whether the local minima 
we obtain are global minima. However, the parameter space of this problem is relatively 
constrained, as the stretched K\"ahler cone is quite narrow. In particular, in order to find a significant 
deviation in the volumes one must go deep into the stretched K\"ahler cone, which is far away from the expected 
location of the minimum (in fact, we always find the minimum to lie on the boundary of the stretched K\"ahler cone). 
In addition, for $h^{1,1} \leq 5$, we attempted the minimization procedure with several data points, 
and with all numerical minimization algorithms available in \textbf{Mathematica}, and always found 
the same minima. We therefore expect the general pattern of minimal volumes to persist. 

Our implicit assumption in computing a single triangulation per polytope is that the anti-canonical volume $\tau$ will not very greatly under changing triangulations. We have verified this by checking all triangulations for polytopes with $h^{1,1} \leq 6$, and found that  $ \mathrm{min}(\mathrm{vol}( \KB ))$ varied by a factor of at most 2.7, discussed further below.

\begin{figure}
\centering
    \textbf{\large$ \mathrm{log}_{10}( \mathrm{min}(\mathrm{vol}( \KB )) )$}\par\medskip
    \includegraphics[width=7cm]{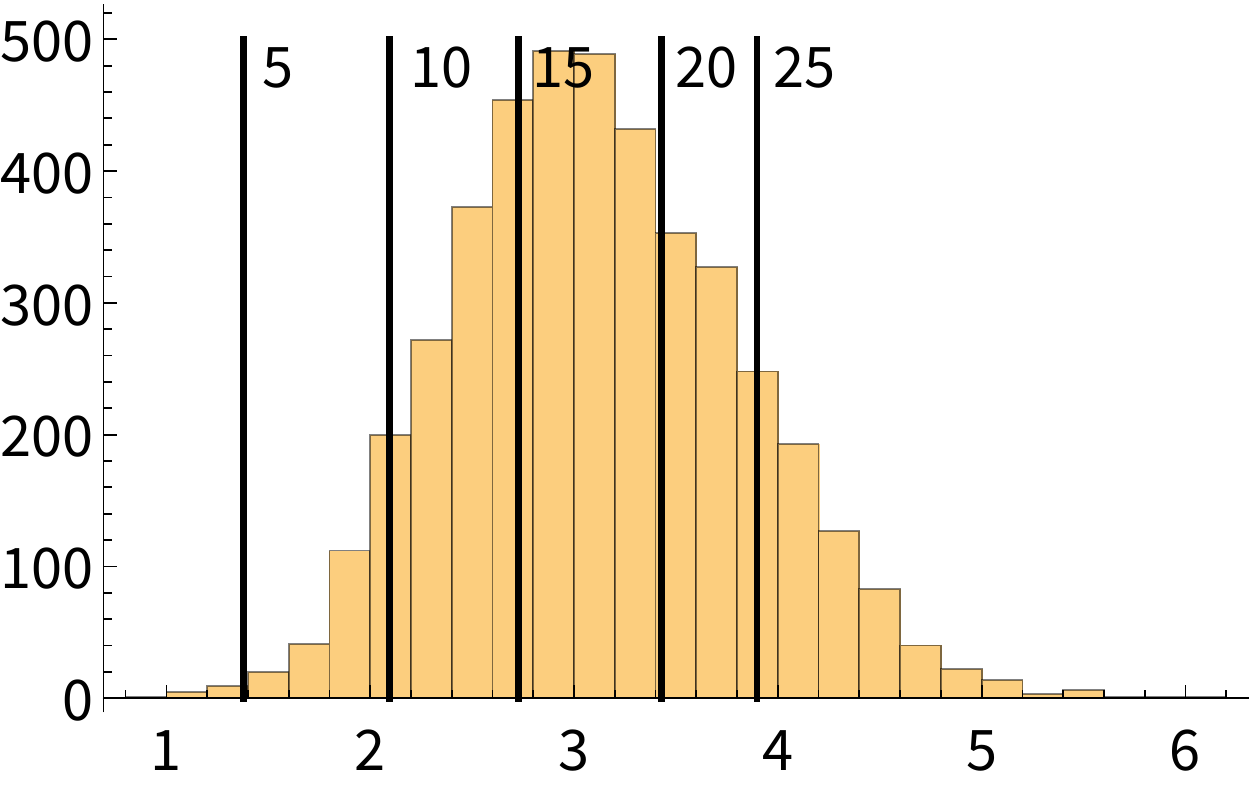}
    \caption{The distribution of $\mathrm{log}_{10}(\mathrm{min}(\mathrm{vol}(\KB)))$ for a single triangulation of each 3d reflexive polytope. The distribution peaks around $\mathrm{vol}(\KB) \sim 10^3$, for which the associated UV gauge coupling is much smaller than the necessary Standard Model couplings. The vertical lines indicate the value of $h^{1,1}$ for which all geometries with that or larger $h^{1,1}$ have minimum anti-canonical volumes to the right of the line.}    \label{fig:1}
 \end{figure}

The distribution of couplings over the entire ensemble is shown in Fig.~\ref{fig:1}. The vertical lines indicate the value of $h^{1,1}$ for which all geometries with that or larger $h^{1,1}$ have minimum anti-canonical volumes to the right of the line. We find 15 polytopes that satisfy $\tau \leq 25$, with a maximal $h^{1,1}$ of five. 
To contrast, let us compare this to the geometries corresponding to the two maximal 3d reflexive polytopes (including $\Diamond_8$) with $h^{1,1} = 35$. 
For a single triangulation of each we find that $\mathrm{min}(\mathrm{vol}(\KB))\ \simeq 14,000$ and $28,000$, respectively. Therefore, to realize a realistic gauge coupling in these examples, $\tau$ would need to be reduced by a factor of about $1000$. Scaling the K\"ahler form $J$ homogeneously $J \rightarrow \lambda J$, we would need to take $\lambda = 1/\sqrt{1000} \simeq 0.03$, resulting in many very small curves and divisors, taking us well outside of the stretched K\"ahler cone, and therefore well outside the region of control. One might hope to find a point in the stretched K\"ahler cone where $\tau$ is greatly reduced; however, due to the narrow nature of the stretched K\"ahler cone~\cite{Demirtas:2018akl} this seems quite unlikely. Finally, there is the question of how much $\tau$ can vary under changing the triangulation. At $h^{1,1} \leq 6$ above we found a difference of at most a factor of 2.7. While we expect the allowed variance to grow with $h^{1,1}$, to get $\tau \leq 50$ the largest toric divisor must have volume at most $12$ (assuming the rest have unit volume), which we expect cannot happen in the stretched K\"ahler cone at even moderate $h^{1,1}$.

Let us finally discuss the statistics of this study in the context of the $\mathbb{P}_{F_{11}}$ models of \cite{Cvetic:2019gnh}. 
We found 15 polytopes that satisfy $\tau \leq 50$, with a maximal $h^{1,1}$ of five. The polytopes with $h^{1,1} \leq 5$ admit 4530 regular fine triangulations. On the other hand, it is expected that the large polytopes, with unrealistic gauge couplings in the stretched K\"ahler cone, admit $\sim \mathcal{O}(10^{15})$ FRSTs, and so we expect a fraction of $\sim 10^{-11}$ of pure MSSM constructions  in this ensemble to admit the correct gauge coupling in a regime of control.

\section{Discussion}\label{sec:discussion}

In this note we have argued for the incompatibility of realistic F-theory compactifications with the following three criterion simultaneously: (i) moderate-to-large $h^{1,1}$, (ii) control of the effective field theory, and (iii) no gauged dark sectors from seven-branes.
The tension arises from reproducing the correct UV-values of the Standard Model's gauge couplings while suppressing certain non-perturbative instanton corrections to the effective theory, in particular the K\"ahler potential, while also satisfying seven-brane tadpole cancellation.

Using the landscape of toric weak Fano threefolds, we have exemplified how for bases with large $h^{1,1}$, tuning the visible sector (be it an MSSM or a GUT scenario) on small divisors generically leads to a large hidden sector.
It is worth noting that there are sometimes MSSM-charged vector-like messengers to these sectors. The extent to which the beyond-MSSM gauge sector is hidden
depends crucially on the messenger mass.

In this landscape there is also a particular setting, namely by tuning the visible sector on anti-canonical divisors, which guarantees the absence of any additional seven-brane gauge sectors.
The recently found ensemble of $\mathcal{O}(10^{15})$ three-family MSSM compactifications \cite{Cvetic:2019gnh} falls into this category.
For these and similar compactifications, we have then analyzed the volumes of the anti-canonical class $\KB$ within the stretched K\"ahler cone.
Minimizing $\text{vol}(\KB) \equiv \tau_{\KB}$ within this part of the K\"ahler moduli space gives an upper bound on the gauge coupling $g_\text{UV}$, via Eq.~\ref{eqn:gvol}, in the supergravity regime.
We find that with this restriction, only around 4500 topologically distinct weak Fano toric threefolds admit realistic gauge couplings, $\alpha_{3,2,1} \simeq 1/25$ for seven-branes on $\KB$. All have $h^{1,1} \leq 5$, which only makes about a fraction of $\sim 10^{-11}$ of the whole weak Fano toric landscape.

\medskip
Avoiding one or more of the three conditions appears necessary and opens the model building possibilities at the expense of encountering other challenges.

Abandoning condition (i) allows for the models of \cite{Cvetic:2019gnh} with the
exact chiral spectrum of the Standard Model to obtain the correct values of the (by construction unified) gauge couplings in a regime of control. 
In fact, $10^4$ of the models allow for this possibility, which is nevertheless
a small fraction of that ensemble. More
generally, given the great deal of evidence suggesting that most string vacua lie at $h^{1,1} > 5$, a large yet-unknown measure factor
would be required to select these vacua. An interesting 
consequence, however, is that these vacua would face fewer
constraints (and discovery possibilities) from axion-like particles.

Abandoning condition (ii) means to give up well-studied moduli stabilization scenarios such as KKLT or LVS, and more generally control of the effective theory.
Such an endeavor is certainly the most challenging, but perhaps also the most exciting direction for future works.
Developing techniques to control towers of worldsheet and ED3-instanton corrections would not only enhance the string model building toolkit, but also allow for more general tests of recently discussed swampland conjectures, such as the swampland distance conjecture \cite{Ooguri:2006in, *Grimm:2018ohb, *Lee:2019wij}, in the context of 4d ${\cal N}=1$ string compactifications.

Lastly, let us entertain the possibility of having a large hidden sector by abandoning assumption (iii).
The main advantage then is that we can put the visible sector on small divisors that give realistic UV gauge couplings inside the stretched K\"ahler cone.
Such scenarios will arguably be much more numerous compared to the models where the only gauge sectors are on anti-canonical divisors.
Turning the argument around, our analysis also implies that within the string landscape there is only a small fraction of models with a controlled effective description that realize just the MSSM.
Instead, the vast majority will have a significant hidden sector, whose phenomenological impact will depend on details, such as consistent $G_4$-background, that remains to be studied.

In summary, we see that while each of the three conditions are appealing by themselves from a phenomenological or model building perspective, their incompatibility with our observed vacuum requires new techniques and ideas to advance toward realistic globally consistent string compactifications in the bulk of the landscape.

\vspace{.2cm}
\noindent{\bf Acknowledgements.} We thank Seung-Joo Lee, Liam McAllister, and Cumrun Vafa for helpful discussions.
The work of M.C.~is supported in part by the DOE (HEP) Award DE-SC0013528, the Fay R.~and Eugene L.~Langberg Endowed Chair and the Slovenian Research Agency (ARRS No.~P1-0306).
J.H.~is supported by NSF CAREER grant PHY-1848089. C.L.~is supported in part by NSF
grant PHY-1719877 and NSF CAREER grant PHY-1848089.

\bibliography{refs}

\begin{thebibliography}{56}%
\makeatletter
\providecommand \@ifxundefined [1]{%
 \@ifx{#1\undefined}
}%
\providecommand \@ifnum [1]{%
 \ifnum #1\expandafter \@firstoftwo
 \else \expandafter \@secondoftwo
 \fi
}%
\providecommand \@ifx [1]{%
 \ifx #1\expandafter \@firstoftwo
 \else \expandafter \@secondoftwo
 \fi
}%
\providecommand \natexlab [1]{#1}%
\providecommand \enquote  [1]{``#1''}%
\providecommand \bibnamefont  [1]{#1}%
\providecommand \bibfnamefont [1]{#1}%
\providecommand \citenamefont [1]{#1}%
\providecommand \href@noop [0]{\@secondoftwo}%
\providecommand \href [0]{\begingroup \@sanitize@url \@href}%
\providecommand \@href[1]{\@@startlink{#1}\@@href}%
\providecommand \@@href[1]{\endgroup#1\@@endlink}%
\providecommand \@sanitize@url [0]{\catcode `\\12\catcode `\$12\catcode
  `\&12\catcode `\#12\catcode `\^12\catcode `\_12\catcode `\%12\relax}%
\providecommand \@@startlink[1]{}%
\providecommand \@@endlink[0]{}%
\providecommand \url  [0]{\begingroup\@sanitize@url \@url }%
\providecommand \@url [1]{\endgroup\@href {#1}{\urlprefix }}%
\providecommand \urlprefix  [0]{URL }%
\providecommand \Eprint [0]{\href }%
\providecommand \doibase [0]{http://dx.doi.org/}%
\providecommand \selectlanguage [0]{\@gobble}%
\providecommand \bibinfo  [0]{\@secondoftwo}%
\providecommand \bibfield  [0]{\@secondoftwo}%
\providecommand \translation [1]{[#1]}%
\providecommand \BibitemOpen [0]{}%
\providecommand \bibitemStop [0]{}%
\providecommand \bibitemNoStop [0]{.\EOS\space}%
\providecommand \EOS [0]{\spacefactor3000\relax}%
\providecommand \BibitemShut  [1]{\csname bibitem#1\endcsname}%
\let\auto@bib@innerbib\@empty
\bibitem [{\citenamefont {Cveti{\v c}}\ \emph
  {et~al.}(2001{\natexlab{a}})\citenamefont {Cveti{\v c}}, \citenamefont
  {Shiu},\ and\ \citenamefont {Uranga}}]{Cvetic:2001nr}%
  \BibitemOpen
  \bibfield  {author} {\bibinfo {author} {\bibfnamefont {M.}~\bibnamefont
  {Cveti{\v c}}}, \bibinfo {author} {\bibfnamefont {G.}~\bibnamefont {Shiu}}, \
  and\ \bibinfo {author} {\bibfnamefont {A.~M.}\ \bibnamefont {Uranga}},\
  }\href {\doibase 10.1016/S0550-3213(01)00427-8} {\bibfield  {journal}
  {\bibinfo  {journal} {Nucl. Phys.}\ }\textbf {\bibinfo {volume} {B615}},\
  \bibinfo {pages} {3} (\bibinfo {year} {2001}{\natexlab{a}})},\ \Eprint
  {http://arxiv.org/abs/hep-th/0107166} {arXiv:hep-th/0107166 [hep-th]}
  \BibitemShut {NoStop}%
\bibitem [{\citenamefont {Cveti{\v c}}\ \emph
  {et~al.}(2001{\natexlab{b}})\citenamefont {Cveti{\v c}}, \citenamefont
  {Shiu},\ and\ \citenamefont {Uranga}}]{Cvetic:2001tj}%
  \BibitemOpen
  \bibfield  {author} {\bibinfo {author} {\bibfnamefont {M.}~\bibnamefont
  {Cveti{\v c}}}, \bibinfo {author} {\bibfnamefont {G.}~\bibnamefont {Shiu}}, \
  and\ \bibinfo {author} {\bibfnamefont {A.~M.}\ \bibnamefont {Uranga}},\
  }\href {\doibase 10.1103/PhysRevLett.87.201801} {\bibfield  {journal}
  {\bibinfo  {journal} {Phys. Rev. Lett.}\ }\textbf {\bibinfo {volume} {87}},\
  \bibinfo {pages} {201801} (\bibinfo {year} {2001}{\natexlab{b}})},\ \Eprint
  {http://arxiv.org/abs/hep-th/0107143} {arXiv:hep-th/0107143 [hep-th]}
  \BibitemShut {NoStop}%
\bibitem [{\citenamefont {Braun}\ \emph {et~al.}(2005)\citenamefont {Braun},
  \citenamefont {He}, \citenamefont {Ovrut},\ and\ \citenamefont
  {Pantev}}]{Braun:2005ux}%
  \BibitemOpen
  \bibfield  {author} {\bibinfo {author} {\bibfnamefont {V.}~\bibnamefont
  {Braun}}, \bibinfo {author} {\bibfnamefont {Y.-H.}\ \bibnamefont {He}},
  \bibinfo {author} {\bibfnamefont {B.~A.}\ \bibnamefont {Ovrut}}, \ and\
  \bibinfo {author} {\bibfnamefont {T.}~\bibnamefont {Pantev}},\ }\href
  {\doibase 10.1016/j.physletb.2005.05.007} {\bibfield  {journal} {\bibinfo
  {journal} {Phys. Lett.}\ }\textbf {\bibinfo {volume} {B618}},\ \bibinfo
  {pages} {252} (\bibinfo {year} {2005})},\ \Eprint
  {http://arxiv.org/abs/hep-th/0501070} {arXiv:hep-th/0501070 [hep-th]}
  \BibitemShut {NoStop}%
\bibitem [{\citenamefont {Bouchard}\ and\ \citenamefont
  {Donagi}(2006)}]{Bouchard:2005ag}%
  \BibitemOpen
  \bibfield  {author} {\bibinfo {author} {\bibfnamefont {V.}~\bibnamefont
  {Bouchard}}\ and\ \bibinfo {author} {\bibfnamefont {R.}~\bibnamefont
  {Donagi}},\ }\href {\doibase 10.1016/j.physletb.2005.12.042} {\bibfield
  {journal} {\bibinfo  {journal} {Phys. Lett.}\ }\textbf {\bibinfo {volume}
  {B633}},\ \bibinfo {pages} {783} (\bibinfo {year} {2006})},\ \Eprint
  {http://arxiv.org/abs/hep-th/0512149} {arXiv:hep-th/0512149 [hep-th]}
  \BibitemShut {NoStop}%
\bibitem [{\citenamefont {Braun}\ \emph {et~al.}(2006)\citenamefont {Braun},
  \citenamefont {He}, \citenamefont {Ovrut},\ and\ \citenamefont
  {Pantev}}]{Braun:2005nv}%
  \BibitemOpen
  \bibfield  {author} {\bibinfo {author} {\bibfnamefont {V.}~\bibnamefont
  {Braun}}, \bibinfo {author} {\bibfnamefont {Y.-H.}\ \bibnamefont {He}},
  \bibinfo {author} {\bibfnamefont {B.~A.}\ \bibnamefont {Ovrut}}, \ and\
  \bibinfo {author} {\bibfnamefont {T.}~\bibnamefont {Pantev}},\ }\href
  {\doibase 10.1088/1126-6708/2006/05/043} {\bibfield  {journal} {\bibinfo
  {journal} {JHEP}\ }\textbf {\bibinfo {volume} {05}},\ \bibinfo {pages} {043}
  (\bibinfo {year} {2006})},\ \Eprint {http://arxiv.org/abs/hep-th/0512177}
  {arXiv:hep-th/0512177 [hep-th]} \BibitemShut {NoStop}%
\bibitem [{\citenamefont {Bouchard}\ \emph {et~al.}(2006)\citenamefont
  {Bouchard}, \citenamefont {Cveti{\v c}},\ and\ \citenamefont
  {Donagi}}]{Bouchard:2006dn}%
  \BibitemOpen
  \bibfield  {author} {\bibinfo {author} {\bibfnamefont {V.}~\bibnamefont
  {Bouchard}}, \bibinfo {author} {\bibfnamefont {M.}~\bibnamefont {Cveti{\v
  c}}}, \ and\ \bibinfo {author} {\bibfnamefont {R.}~\bibnamefont {Donagi}},\
  }\href {\doibase 10.1016/j.nuclphysb.2006.03.032} {\bibfield  {journal}
  {\bibinfo  {journal} {Nucl. Phys.}\ }\textbf {\bibinfo {volume} {B745}},\
  \bibinfo {pages} {62} (\bibinfo {year} {2006})},\ \Eprint
  {http://arxiv.org/abs/hep-th/0602096} {arXiv:hep-th/0602096 [hep-th]}
  \BibitemShut {NoStop}%
\bibitem [{\citenamefont {Anderson}\ \emph {et~al.}(2007)\citenamefont
  {Anderson}, \citenamefont {He},\ and\ \citenamefont
  {Lukas}}]{Anderson:2007nc}%
  \BibitemOpen
  \bibfield  {author} {\bibinfo {author} {\bibfnamefont {L.~B.}\ \bibnamefont
  {Anderson}}, \bibinfo {author} {\bibfnamefont {Y.-H.}\ \bibnamefont {He}}, \
  and\ \bibinfo {author} {\bibfnamefont {A.}~\bibnamefont {Lukas}},\ }\href
  {\doibase 10.1088/1126-6708/2007/07/049} {\bibfield  {journal} {\bibinfo
  {journal} {JHEP}\ }\textbf {\bibinfo {volume} {07}},\ \bibinfo {pages} {049}
  (\bibinfo {year} {2007})},\ \Eprint {http://arxiv.org/abs/hep-th/0702210}
  {arXiv:hep-th/0702210 [HEP-TH]} \BibitemShut {NoStop}%
\bibitem [{\citenamefont {Anderson}\ \emph {et~al.}(2010)\citenamefont
  {Anderson}, \citenamefont {Gray}, \citenamefont {He},\ and\ \citenamefont
  {Lukas}}]{Anderson:2009mh}%
  \BibitemOpen
  \bibfield  {author} {\bibinfo {author} {\bibfnamefont {L.~B.}\ \bibnamefont
  {Anderson}}, \bibinfo {author} {\bibfnamefont {J.}~\bibnamefont {Gray}},
  \bibinfo {author} {\bibfnamefont {Y.-H.}\ \bibnamefont {He}}, \ and\ \bibinfo
  {author} {\bibfnamefont {A.}~\bibnamefont {Lukas}},\ }\href {\doibase
  10.1007/JHEP02(2010)054} {\bibfield  {journal} {\bibinfo  {journal} {JHEP}\
  }\textbf {\bibinfo {volume} {02}},\ \bibinfo {pages} {054} (\bibinfo {year}
  {2010})},\ \Eprint {http://arxiv.org/abs/0911.1569} {arXiv:0911.1569
  [hep-th]} \BibitemShut {NoStop}%
\bibitem [{\citenamefont {Cveti{\v c}}\ \emph {et~al.}(2015)\citenamefont
  {Cveti{\v c}}, \citenamefont {Klevers}, \citenamefont {Pe{\~n}a},
  \citenamefont {Oehlmann},\ and\ \citenamefont {Reuter}}]{Cvetic:2015txa}%
  \BibitemOpen
  \bibfield  {author} {\bibinfo {author} {\bibfnamefont {M.}~\bibnamefont
  {Cveti{\v c}}}, \bibinfo {author} {\bibfnamefont {D.}~\bibnamefont
  {Klevers}}, \bibinfo {author} {\bibfnamefont {D.~K.~M.}\ \bibnamefont
  {Pe{\~n}a}}, \bibinfo {author} {\bibfnamefont {P.-K.}\ \bibnamefont
  {Oehlmann}}, \ and\ \bibinfo {author} {\bibfnamefont {J.}~\bibnamefont
  {Reuter}},\ }\href {\doibase 10.1007/JHEP08(2015)087} {\bibfield  {journal}
  {\bibinfo  {journal} {JHEP}\ }\textbf {\bibinfo {volume} {08}},\ \bibinfo
  {pages} {087} (\bibinfo {year} {2015})},\ \Eprint
  {http://arxiv.org/abs/1503.02068} {arXiv:1503.02068 [hep-th]} \BibitemShut
  {NoStop}%
\bibitem [{\citenamefont {Lin}\ and\ \citenamefont
  {Weigand}(2015)}]{Lin:2014qga}%
  \BibitemOpen
  \bibfield  {author} {\bibinfo {author} {\bibfnamefont {L.}~\bibnamefont
  {Lin}}\ and\ \bibinfo {author} {\bibfnamefont {T.}~\bibnamefont {Weigand}},\
  }\href {\doibase 10.1002/prop.201400072} {\bibfield  {journal} {\bibinfo
  {journal} {Fortsch. Phys.}\ }\textbf {\bibinfo {volume} {63}},\ \bibinfo
  {pages} {55} (\bibinfo {year} {2015})},\ \Eprint
  {http://arxiv.org/abs/1406.6071} {arXiv:1406.6071 [hep-th]} \BibitemShut
  {NoStop}%
\bibitem [{\citenamefont {Lin}\ and\ \citenamefont
  {Weigand}(2016)}]{Lin:2016vus}%
  \BibitemOpen
  \bibfield  {author} {\bibinfo {author} {\bibfnamefont {L.}~\bibnamefont
  {Lin}}\ and\ \bibinfo {author} {\bibfnamefont {T.}~\bibnamefont {Weigand}},\
  }\href {\doibase 10.1016/j.nuclphysb.2016.09.008} {\bibfield  {journal}
  {\bibinfo  {journal} {Nucl. Phys.}\ }\textbf {\bibinfo {volume} {B913}},\
  \bibinfo {pages} {209} (\bibinfo {year} {2016})},\ \Eprint
  {http://arxiv.org/abs/1604.04292} {arXiv:1604.04292 [hep-th]} \BibitemShut
  {NoStop}%
\bibitem [{\citenamefont {Cveti{\v c}}\ \emph {et~al.}(2018)\citenamefont
  {Cveti{\v c}}, \citenamefont {Lin}, \citenamefont {Liu},\ and\ \citenamefont
  {Oehlmann}}]{Cvetic:2018ryq}%
  \BibitemOpen
  \bibfield  {author} {\bibinfo {author} {\bibfnamefont {M.}~\bibnamefont
  {Cveti{\v c}}}, \bibinfo {author} {\bibfnamefont {L.}~\bibnamefont {Lin}},
  \bibinfo {author} {\bibfnamefont {M.}~\bibnamefont {Liu}}, \ and\ \bibinfo
  {author} {\bibfnamefont {P.-K.}\ \bibnamefont {Oehlmann}},\ }\href {\doibase
  10.1007/JHEP09(2018)089} {\bibfield  {journal} {\bibinfo  {journal} {JHEP}\
  }\textbf {\bibinfo {volume} {09}},\ \bibinfo {pages} {089} (\bibinfo {year}
  {2018})},\ \Eprint {http://arxiv.org/abs/1807.01320} {arXiv:1807.01320
  [hep-th]} \BibitemShut {NoStop}%
\bibitem [{\citenamefont {Cveti{\v c}}\ \emph {et~al.}(2019)\citenamefont
  {Cveti{\v c}}, \citenamefont {Halverson}, \citenamefont {Lin}, \citenamefont
  {Liu},\ and\ \citenamefont {Tian}}]{Cvetic:2019gnh}%
  \BibitemOpen
  \bibfield  {author} {\bibinfo {author} {\bibfnamefont {M.}~\bibnamefont
  {Cveti{\v c}}}, \bibinfo {author} {\bibfnamefont {J.}~\bibnamefont
  {Halverson}}, \bibinfo {author} {\bibfnamefont {L.}~\bibnamefont {Lin}},
  \bibinfo {author} {\bibfnamefont {M.}~\bibnamefont {Liu}}, \ and\ \bibinfo
  {author} {\bibfnamefont {J.}~\bibnamefont {Tian}},\ }\href {\doibase
  10.1103/PhysRevLett.123.101601} {\bibfield  {journal} {\bibinfo  {journal}
  {Phys. Rev. Lett.}\ }\textbf {\bibinfo {volume} {123}},\ \bibinfo {pages}
  {101601} (\bibinfo {year} {2019})},\ \Eprint
  {http://arxiv.org/abs/1903.00009} {arXiv:1903.00009 [hep-th]} \BibitemShut
  {NoStop}%
\bibitem [{\citenamefont {Grassi}\ \emph {et~al.}(2015)\citenamefont {Grassi},
  \citenamefont {Halverson}, \citenamefont {Shaneson},\ and\ \citenamefont
  {Taylor}}]{Grassi:2014zxa}%
  \BibitemOpen
  \bibfield  {author} {\bibinfo {author} {\bibfnamefont {A.}~\bibnamefont
  {Grassi}}, \bibinfo {author} {\bibfnamefont {J.}~\bibnamefont {Halverson}},
  \bibinfo {author} {\bibfnamefont {J.}~\bibnamefont {Shaneson}}, \ and\
  \bibinfo {author} {\bibfnamefont {W.}~\bibnamefont {Taylor}},\ }\href
  {\doibase 10.1007/JHEP01(2015)086} {\bibfield  {journal} {\bibinfo  {journal}
  {JHEP}\ }\textbf {\bibinfo {volume} {01}},\ \bibinfo {pages} {086} (\bibinfo
  {year} {2015})},\ \Eprint {http://arxiv.org/abs/1409.8295} {arXiv:1409.8295
  [hep-th]} \BibitemShut {NoStop}%
\bibitem [{\citenamefont {Morrison}\ and\ \citenamefont
  {Taylor}(2015)}]{Morrison:2014lca}%
  \BibitemOpen
  \bibfield  {author} {\bibinfo {author} {\bibfnamefont {D.~R.}\ \bibnamefont
  {Morrison}}\ and\ \bibinfo {author} {\bibfnamefont {W.}~\bibnamefont
  {Taylor}},\ }\href {\doibase 10.1007/JHEP05(2015)080} {\bibfield  {journal}
  {\bibinfo  {journal} {JHEP}\ }\textbf {\bibinfo {volume} {05}},\ \bibinfo
  {pages} {080} (\bibinfo {year} {2015})},\ \Eprint
  {http://arxiv.org/abs/1412.6112} {arXiv:1412.6112 [hep-th]} \BibitemShut
  {NoStop}%
\bibitem [{\citenamefont {Halverson}\ and\ \citenamefont
  {Taylor}(2015)}]{Halverson:2015jua}%
  \BibitemOpen
  \bibfield  {author} {\bibinfo {author} {\bibfnamefont {J.}~\bibnamefont
  {Halverson}}\ and\ \bibinfo {author} {\bibfnamefont {W.}~\bibnamefont
  {Taylor}},\ }\href {\doibase 10.1007/JHEP09(2015)086} {\bibfield  {journal}
  {\bibinfo  {journal} {JHEP}\ }\textbf {\bibinfo {volume} {09}},\ \bibinfo
  {pages} {086} (\bibinfo {year} {2015})},\ \Eprint
  {http://arxiv.org/abs/1506.03204} {arXiv:1506.03204 [hep-th]} \BibitemShut
  {NoStop}%
\bibitem [{\citenamefont {Taylor}\ and\ \citenamefont
  {Wang}(2016)}]{Taylor:2015ppa}%
  \BibitemOpen
  \bibfield  {author} {\bibinfo {author} {\bibfnamefont {W.}~\bibnamefont
  {Taylor}}\ and\ \bibinfo {author} {\bibfnamefont {Y.-N.}\ \bibnamefont
  {Wang}},\ }\href {\doibase 10.1007/JHEP01(2016)137} {\bibfield  {journal}
  {\bibinfo  {journal} {JHEP}\ }\textbf {\bibinfo {volume} {01}},\ \bibinfo
  {pages} {137} (\bibinfo {year} {2016})},\ \Eprint
  {http://arxiv.org/abs/1510.04978} {arXiv:1510.04978 [hep-th]} \BibitemShut
  {NoStop}%
\bibitem [{\citenamefont {Taylor}\ and\ \citenamefont
  {Wang}(2015)}]{Taylor:2015xtz}%
  \BibitemOpen
  \bibfield  {author} {\bibinfo {author} {\bibfnamefont {W.}~\bibnamefont
  {Taylor}}\ and\ \bibinfo {author} {\bibfnamefont {Y.-N.}\ \bibnamefont
  {Wang}},\ }\href {\doibase 10.1007/JHEP12(2015)164} {\bibfield  {journal}
  {\bibinfo  {journal} {JHEP}\ }\textbf {\bibinfo {volume} {12}},\ \bibinfo
  {pages} {164} (\bibinfo {year} {2015})},\ \Eprint
  {http://arxiv.org/abs/1511.03209} {arXiv:1511.03209 [hep-th]} \BibitemShut
  {NoStop}%
\bibitem [{\citenamefont {Taylor}\ and\ \citenamefont
  {Wang}(2018)}]{Taylor:2017yqr}%
  \BibitemOpen
  \bibfield  {author} {\bibinfo {author} {\bibfnamefont {W.}~\bibnamefont
  {Taylor}}\ and\ \bibinfo {author} {\bibfnamefont {Y.-N.}\ \bibnamefont
  {Wang}},\ }\href {\doibase 10.1007/JHEP01(2018)111} {\bibfield  {journal}
  {\bibinfo  {journal} {JHEP}\ }\textbf {\bibinfo {volume} {01}},\ \bibinfo
  {pages} {111} (\bibinfo {year} {2018})},\ \Eprint
  {http://arxiv.org/abs/1710.11235} {arXiv:1710.11235 [hep-th]} \BibitemShut
  {NoStop}%
\bibitem [{\citenamefont {Halverson}\ \emph
  {et~al.}(2017{\natexlab{a}})\citenamefont {Halverson}, \citenamefont {Long},\
  and\ \citenamefont {Sung}}]{Halverson:2017ffz}%
  \BibitemOpen
  \bibfield  {author} {\bibinfo {author} {\bibfnamefont {J.}~\bibnamefont
  {Halverson}}, \bibinfo {author} {\bibfnamefont {C.}~\bibnamefont {Long}}, \
  and\ \bibinfo {author} {\bibfnamefont {B.}~\bibnamefont {Sung}},\ }\href
  {\doibase 10.1103/PhysRevD.96.126006} {\bibfield  {journal} {\bibinfo
  {journal} {Phys. Rev.}\ }\textbf {\bibinfo {volume} {D96}},\ \bibinfo {pages}
  {126006} (\bibinfo {year} {2017}{\natexlab{a}})},\ \Eprint
  {http://arxiv.org/abs/1706.02299} {arXiv:1706.02299 [hep-th]} \BibitemShut
  {NoStop}%
\bibitem [{\citenamefont {Halverson}\ \emph {et~al.}(2018)\citenamefont
  {Halverson}, \citenamefont {Long},\ and\ \citenamefont
  {Sung}}]{Halverson:2017vde}%
  \BibitemOpen
  \bibfield  {author} {\bibinfo {author} {\bibfnamefont {J.}~\bibnamefont
  {Halverson}}, \bibinfo {author} {\bibfnamefont {C.}~\bibnamefont {Long}}, \
  and\ \bibinfo {author} {\bibfnamefont {B.}~\bibnamefont {Sung}},\ }\href
  {\doibase 10.1007/JHEP02(2018)113} {\bibfield  {journal} {\bibinfo  {journal}
  {JHEP}\ }\textbf {\bibinfo {volume} {02}},\ \bibinfo {pages} {113} (\bibinfo
  {year} {2018})},\ \Eprint {http://arxiv.org/abs/1710.09374} {arXiv:1710.09374
  [hep-th]} \BibitemShut {NoStop}%
\bibitem [{\citenamefont {Wang}(2020)}]{Wang:2020gmi}%
  \BibitemOpen
  \bibfield  {author} {\bibinfo {author} {\bibfnamefont {Y.-N.}\ \bibnamefont
  {Wang}},\ }\href@noop {} {\  (\bibinfo {year} {2020})},\ \Eprint
  {http://arxiv.org/abs/2001.07258} {arXiv:2001.07258 [hep-th]} \BibitemShut
  {NoStop}%
\bibitem [{\citenamefont {Kachru}\ \emph {et~al.}(2003)\citenamefont {Kachru},
  \citenamefont {Kallosh}, \citenamefont {Linde},\ and\ \citenamefont
  {Trivedi}}]{Kachru:2003aw}%
  \BibitemOpen
  \bibfield  {author} {\bibinfo {author} {\bibfnamefont {S.}~\bibnamefont
  {Kachru}}, \bibinfo {author} {\bibfnamefont {R.}~\bibnamefont {Kallosh}},
  \bibinfo {author} {\bibfnamefont {A.~D.}\ \bibnamefont {Linde}}, \ and\
  \bibinfo {author} {\bibfnamefont {S.~P.}\ \bibnamefont {Trivedi}},\ }\href
  {\doibase 10.1103/PhysRevD.68.046005} {\bibfield  {journal} {\bibinfo
  {journal} {Phys. Rev.}\ }\textbf {\bibinfo {volume} {D68}},\ \bibinfo {pages}
  {046005} (\bibinfo {year} {2003})},\ \Eprint
  {http://arxiv.org/abs/hep-th/0301240} {arXiv:hep-th/0301240 [hep-th]}
  \BibitemShut {NoStop}%
\bibitem [{\citenamefont {Balasubramanian}\ \emph {et~al.}(2005)\citenamefont
  {Balasubramanian}, \citenamefont {Berglund}, \citenamefont {Conlon},\ and\
  \citenamefont {Quevedo}}]{Balasubramanian:2005zx}%
  \BibitemOpen
  \bibfield  {author} {\bibinfo {author} {\bibfnamefont {V.}~\bibnamefont
  {Balasubramanian}}, \bibinfo {author} {\bibfnamefont {P.}~\bibnamefont
  {Berglund}}, \bibinfo {author} {\bibfnamefont {J.~P.}\ \bibnamefont
  {Conlon}}, \ and\ \bibinfo {author} {\bibfnamefont {F.}~\bibnamefont
  {Quevedo}},\ }\href {\doibase 10.1088/1126-6708/2005/03/007} {\bibfield
  {journal} {\bibinfo  {journal} {JHEP}\ }\textbf {\bibinfo {volume} {03}},\
  \bibinfo {pages} {007} (\bibinfo {year} {2005})},\ \Eprint
  {http://arxiv.org/abs/hep-th/0502058} {arXiv:hep-th/0502058 [hep-th]}
  \BibitemShut {NoStop}%
\bibitem [{\citenamefont {Demirtas}\ \emph {et~al.}(2018)\citenamefont
  {Demirtas}, \citenamefont {Long}, \citenamefont {McAllister},\ and\
  \citenamefont {Stillman}}]{Demirtas:2018akl}%
  \BibitemOpen
  \bibfield  {author} {\bibinfo {author} {\bibfnamefont {M.}~\bibnamefont
  {Demirtas}}, \bibinfo {author} {\bibfnamefont {C.}~\bibnamefont {Long}},
  \bibinfo {author} {\bibfnamefont {L.}~\bibnamefont {McAllister}}, \ and\
  \bibinfo {author} {\bibfnamefont {M.}~\bibnamefont {Stillman}},\ }\href@noop
  {} {\  (\bibinfo {year} {2018})},\ \Eprint {http://arxiv.org/abs/1808.01282}
  {arXiv:1808.01282 [hep-th]} \BibitemShut {NoStop}%
\bibitem [{\citenamefont {Halverson}\ \emph
  {et~al.}(2019{\natexlab{a}})\citenamefont {Halverson}, \citenamefont {Long},
  \citenamefont {Nelson},\ and\ \citenamefont {Salinas}}]{Halverson:2019cmy}%
  \BibitemOpen
  \bibfield  {author} {\bibinfo {author} {\bibfnamefont {J.}~\bibnamefont
  {Halverson}}, \bibinfo {author} {\bibfnamefont {C.}~\bibnamefont {Long}},
  \bibinfo {author} {\bibfnamefont {B.}~\bibnamefont {Nelson}}, \ and\ \bibinfo
  {author} {\bibfnamefont {G.}~\bibnamefont {Salinas}},\ }\href {\doibase
  10.1103/PhysRevD.100.106010} {\bibfield  {journal} {\bibinfo  {journal}
  {Phys. Rev.}\ }\textbf {\bibinfo {volume} {D100}},\ \bibinfo {pages} {106010}
  (\bibinfo {year} {2019}{\natexlab{a}})},\ \Eprint
  {http://arxiv.org/abs/1909.05257} {arXiv:1909.05257 [hep-th]} \BibitemShut
  {NoStop}%
\bibitem [{\citenamefont {Weigand}(2018)}]{Weigand:2018rez}%
  \BibitemOpen
  \bibfield  {author} {\bibinfo {author} {\bibfnamefont {T.}~\bibnamefont
  {Weigand}},\ }\href@noop {} {\bibfield  {journal} {\bibinfo  {journal} {PoS}\
  }\textbf {\bibinfo {volume} {TASI2017}},\ \bibinfo {pages} {016} (\bibinfo
  {year} {2018})},\ \Eprint {http://arxiv.org/abs/1806.01854} {arXiv:1806.01854
  [hep-th]} \BibitemShut {NoStop}%
\bibitem [{\citenamefont {Cveti{\v c}}\ and\ \citenamefont
  {Lin}(2018{\natexlab{a}})}]{Cvetic:2018bni}%
  \BibitemOpen
  \bibfield  {author} {\bibinfo {author} {\bibfnamefont {M.}~\bibnamefont
  {Cveti{\v c}}}\ and\ \bibinfo {author} {\bibfnamefont {L.}~\bibnamefont
  {Lin}},\ }\href {\doibase 10.22323/1.305.0020} {\bibfield  {journal}
  {\bibinfo  {journal} {PoS}\ }\textbf {\bibinfo {volume} {TASI2017}},\
  \bibinfo {pages} {020} (\bibinfo {year} {2018}{\natexlab{a}})},\ \Eprint
  {http://arxiv.org/abs/1809.00012} {arXiv:1809.00012 [hep-th]} \BibitemShut
  {NoStop}%
\bibitem [{\citenamefont {Grimm}(2011)}]{Grimm:2010ks}%
  \BibitemOpen
  \bibfield  {author} {\bibinfo {author} {\bibfnamefont {T.~W.}\ \bibnamefont
  {Grimm}},\ }\href {\doibase 10.1016/j.nuclphysb.2010.11.018} {\bibfield
  {journal} {\bibinfo  {journal} {Nucl. Phys.}\ }\textbf {\bibinfo {volume}
  {B845}},\ \bibinfo {pages} {48} (\bibinfo {year} {2011})},\ \Eprint
  {http://arxiv.org/abs/1008.4133} {arXiv:1008.4133 [hep-th]} \BibitemShut
  {NoStop}%
\bibitem [{\citenamefont {Hosono}\ \emph {et~al.}(1994)\citenamefont {Hosono},
  \citenamefont {Klemm},\ and\ \citenamefont {Theisen}}]{Hosono:1994av}%
  \BibitemOpen
  \bibfield  {author} {\bibinfo {author} {\bibfnamefont {S.}~\bibnamefont
  {Hosono}}, \bibinfo {author} {\bibfnamefont {A.}~\bibnamefont {Klemm}}, \
  and\ \bibinfo {author} {\bibfnamefont {S.}~\bibnamefont {Theisen}},\
  }\bibfield  {booktitle} {\emph {\bibinfo {booktitle} {{Proceedings, 3rd
  Baltic Rim Student Seminar: Integrable Models and Strings: Helsinki, Finland,
  September 13-17, 1993}}},\ }\href {\doibase 10.1007/3-540-58453-6_13}
  {\bibfield  {journal} {\bibinfo  {journal} {Lect. Notes Phys.}\ }\textbf
  {\bibinfo {volume} {436}},\ \bibinfo {pages} {235} (\bibinfo {year}
  {1994})},\ \Eprint {http://arxiv.org/abs/hep-th/9403096}
  {arXiv:hep-th/9403096 [hep-th]} \BibitemShut {NoStop}%
\bibitem [{\citenamefont {Candelas}\ \emph {et~al.}(1998)\citenamefont
  {Candelas}, \citenamefont {De~La~Ossa}, \citenamefont {Green},\ and\
  \citenamefont {Parkes}}]{Candelas:1990rm}%
  \BibitemOpen
  \bibfield  {author} {\bibinfo {author} {\bibfnamefont {P.}~\bibnamefont
  {Candelas}}, \bibinfo {author} {\bibfnamefont {X.~C.}\ \bibnamefont
  {De~La~Ossa}}, \bibinfo {author} {\bibfnamefont {P.~S.}\ \bibnamefont
  {Green}}, \ and\ \bibinfo {author} {\bibfnamefont {L.}~\bibnamefont
  {Parkes}},\ }\href {\doibase 10.1016/0550-3213(91)90292-6} {\bibfield
  {journal} {\bibinfo  {journal} {AMS/IP Stud. Adv. Math.}\ }\textbf {\bibinfo
  {volume} {9}},\ \bibinfo {pages} {31} (\bibinfo {year} {1998})}\BibitemShut
  {NoStop}%
\bibitem [{\citenamefont {Demirtas}\ \emph {et~al.}(2019)\citenamefont
  {Demirtas}, \citenamefont {Long}, \citenamefont {McAllister},\ and\
  \citenamefont {Stillman}}]{Demirtas:2019lfi}%
  \BibitemOpen
  \bibfield  {author} {\bibinfo {author} {\bibfnamefont {M.}~\bibnamefont
  {Demirtas}}, \bibinfo {author} {\bibfnamefont {C.}~\bibnamefont {Long}},
  \bibinfo {author} {\bibfnamefont {L.}~\bibnamefont {McAllister}}, \ and\
  \bibinfo {author} {\bibfnamefont {M.}~\bibnamefont {Stillman}},\ }\href@noop
  {} {\  (\bibinfo {year} {2019})},\ \Eprint {http://arxiv.org/abs/1906.08262}
  {arXiv:1906.08262 [hep-th]} \BibitemShut {NoStop}%
\bibitem [{\citenamefont {Halverson}\ \emph
  {et~al.}(2017{\natexlab{b}})\citenamefont {Halverson}, \citenamefont
  {Nelson},\ and\ \citenamefont {Ruehle}}]{Halverson:2016nfq}%
  \BibitemOpen
  \bibfield  {author} {\bibinfo {author} {\bibfnamefont {J.}~\bibnamefont
  {Halverson}}, \bibinfo {author} {\bibfnamefont {B.~D.}\ \bibnamefont
  {Nelson}}, \ and\ \bibinfo {author} {\bibfnamefont {F.}~\bibnamefont
  {Ruehle}},\ }\href {\doibase 10.1103/PhysRevD.95.043527} {\bibfield
  {journal} {\bibinfo  {journal} {Phys. Rev.}\ }\textbf {\bibinfo {volume}
  {D95}},\ \bibinfo {pages} {043527} (\bibinfo {year} {2017}{\natexlab{b}})},\
  \Eprint {http://arxiv.org/abs/1609.02151} {arXiv:1609.02151 [hep-ph]}
  \BibitemShut {NoStop}%
\bibitem [{\citenamefont {Halverson}\ \emph
  {et~al.}(2019{\natexlab{b}})\citenamefont {Halverson}, \citenamefont {Long},
  \citenamefont {Nelson},\ and\ \citenamefont {Salinas}}]{Halverson:2019kna}%
  \BibitemOpen
  \bibfield  {author} {\bibinfo {author} {\bibfnamefont {J.}~\bibnamefont
  {Halverson}}, \bibinfo {author} {\bibfnamefont {C.}~\bibnamefont {Long}},
  \bibinfo {author} {\bibfnamefont {B.}~\bibnamefont {Nelson}}, \ and\ \bibinfo
  {author} {\bibfnamefont {G.}~\bibnamefont {Salinas}},\ }\href {\doibase
  10.1103/PhysRevD.99.086014} {\bibfield  {journal} {\bibinfo  {journal} {Phys.
  Rev.}\ }\textbf {\bibinfo {volume} {D99}},\ \bibinfo {pages} {086014}
  (\bibinfo {year} {2019}{\natexlab{b}})},\ \Eprint
  {http://arxiv.org/abs/1903.04495} {arXiv:1903.04495 [hep-th]} \BibitemShut
  {NoStop}%
\bibitem [{\citenamefont {Carifio}\ \emph {et~al.}(2018)\citenamefont
  {Carifio}, \citenamefont {Cunningham}, \citenamefont {Halverson},
  \citenamefont {Krioukov}, \citenamefont {Long},\ and\ \citenamefont
  {Nelson}}]{Carifio:2017nyb}%
  \BibitemOpen
  \bibfield  {author} {\bibinfo {author} {\bibfnamefont {J.}~\bibnamefont
  {Carifio}}, \bibinfo {author} {\bibfnamefont {W.~J.}\ \bibnamefont
  {Cunningham}}, \bibinfo {author} {\bibfnamefont {J.}~\bibnamefont
  {Halverson}}, \bibinfo {author} {\bibfnamefont {D.}~\bibnamefont {Krioukov}},
  \bibinfo {author} {\bibfnamefont {C.}~\bibnamefont {Long}}, \ and\ \bibinfo
  {author} {\bibfnamefont {B.~D.}\ \bibnamefont {Nelson}},\ }\href {\doibase
  10.1103/PhysRevLett.121.101602} {\bibfield  {journal} {\bibinfo  {journal}
  {Phys. Rev. Lett.}\ }\textbf {\bibinfo {volume} {121}},\ \bibinfo {pages}
  {101602} (\bibinfo {year} {2018})},\ \Eprint
  {http://arxiv.org/abs/1711.06685} {arXiv:1711.06685 [hep-th]} \BibitemShut
  {NoStop}%
\bibitem [{\citenamefont {Cicoli}\ \emph {et~al.}(2012)\citenamefont {Cicoli},
  \citenamefont {Krippendorf}, \citenamefont {Mayrhofer}, \citenamefont
  {Quevedo},\ and\ \citenamefont {Valandro}}]{Cicoli:2012vw}%
  \BibitemOpen
  \bibfield  {author} {\bibinfo {author} {\bibfnamefont {M.}~\bibnamefont
  {Cicoli}}, \bibinfo {author} {\bibfnamefont {S.}~\bibnamefont {Krippendorf}},
  \bibinfo {author} {\bibfnamefont {C.}~\bibnamefont {Mayrhofer}}, \bibinfo
  {author} {\bibfnamefont {F.}~\bibnamefont {Quevedo}}, \ and\ \bibinfo
  {author} {\bibfnamefont {R.}~\bibnamefont {Valandro}},\ }\href {\doibase
  10.1007/JHEP09(2012)019} {\bibfield  {journal} {\bibinfo  {journal} {JHEP}\
  }\textbf {\bibinfo {volume} {09}},\ \bibinfo {pages} {019} (\bibinfo {year}
  {2012})},\ \Eprint {http://arxiv.org/abs/1206.5237} {arXiv:1206.5237
  [hep-th]} \BibitemShut {NoStop}%
\bibitem [{\citenamefont {Cicoli}\ \emph {et~al.}(2013)\citenamefont {Cicoli},
  \citenamefont {Krippendorf}, \citenamefont {Mayrhofer}, \citenamefont
  {Quevedo},\ and\ \citenamefont {Valandro}}]{Cicoli:2013mpa}%
  \BibitemOpen
  \bibfield  {author} {\bibinfo {author} {\bibfnamefont {M.}~\bibnamefont
  {Cicoli}}, \bibinfo {author} {\bibfnamefont {S.}~\bibnamefont {Krippendorf}},
  \bibinfo {author} {\bibfnamefont {C.}~\bibnamefont {Mayrhofer}}, \bibinfo
  {author} {\bibfnamefont {F.}~\bibnamefont {Quevedo}}, \ and\ \bibinfo
  {author} {\bibfnamefont {R.}~\bibnamefont {Valandro}},\ }\href {\doibase
  10.1007/JHEP07(2013)150} {\bibfield  {journal} {\bibinfo  {journal} {JHEP}\
  }\textbf {\bibinfo {volume} {07}},\ \bibinfo {pages} {150} (\bibinfo {year}
  {2013})},\ \Eprint {http://arxiv.org/abs/1304.0022} {arXiv:1304.0022
  [hep-th]} \BibitemShut {NoStop}%
\bibitem [{\citenamefont {Cicoli}\ \emph {et~al.}(2014)\citenamefont {Cicoli},
  \citenamefont {Klevers}, \citenamefont {Krippendorf}, \citenamefont
  {Mayrhofer}, \citenamefont {Quevedo},\ and\ \citenamefont
  {Valandro}}]{Cicoli:2013cha}%
  \BibitemOpen
  \bibfield  {author} {\bibinfo {author} {\bibfnamefont {M.}~\bibnamefont
  {Cicoli}}, \bibinfo {author} {\bibfnamefont {D.}~\bibnamefont {Klevers}},
  \bibinfo {author} {\bibfnamefont {S.}~\bibnamefont {Krippendorf}}, \bibinfo
  {author} {\bibfnamefont {C.}~\bibnamefont {Mayrhofer}}, \bibinfo {author}
  {\bibfnamefont {F.}~\bibnamefont {Quevedo}}, \ and\ \bibinfo {author}
  {\bibfnamefont {R.}~\bibnamefont {Valandro}},\ }\href {\doibase
  10.1007/JHEP05(2014)001} {\bibfield  {journal} {\bibinfo  {journal} {JHEP}\
  }\textbf {\bibinfo {volume} {05}},\ \bibinfo {pages} {001} (\bibinfo {year}
  {2014})},\ \Eprint {http://arxiv.org/abs/1312.0014} {arXiv:1312.0014
  [hep-th]} \BibitemShut {NoStop}%
\bibitem [{\citenamefont {Kreuzer}\ and\ \citenamefont
  {Skarke}(1998)}]{Kreuzer:1998vb}%
  \BibitemOpen
  \bibfield  {author} {\bibinfo {author} {\bibfnamefont {M.}~\bibnamefont
  {Kreuzer}}\ and\ \bibinfo {author} {\bibfnamefont {H.}~\bibnamefont
  {Skarke}},\ }\href {\doibase 10.4310/ATMP.1998.v2.n4.a5} {\bibfield
  {journal} {\bibinfo  {journal} {Adv. Theor. Math. Phys.}\ }\textbf {\bibinfo
  {volume} {2}},\ \bibinfo {pages} {853} (\bibinfo {year} {1998})},\ \Eprint
  {http://arxiv.org/abs/hep-th/9805190} {arXiv:hep-th/9805190 [hep-th]}
  \BibitemShut {NoStop}%
\bibitem [{\citenamefont {Braun}\ \emph {et~al.}(2017)\citenamefont {Braun},
  \citenamefont {Long}, \citenamefont {McAllister}, \citenamefont {Stillman},\
  and\ \citenamefont {Sung}}]{Braun:2017nhi}%
  \BibitemOpen
  \bibfield  {author} {\bibinfo {author} {\bibfnamefont {A.~P.}\ \bibnamefont
  {Braun}}, \bibinfo {author} {\bibfnamefont {C.}~\bibnamefont {Long}},
  \bibinfo {author} {\bibfnamefont {L.}~\bibnamefont {McAllister}}, \bibinfo
  {author} {\bibfnamefont {M.}~\bibnamefont {Stillman}}, \ and\ \bibinfo
  {author} {\bibfnamefont {B.}~\bibnamefont {Sung}},\ }\href@noop {} {\
  (\bibinfo {year} {2017})},\ \Eprint {http://arxiv.org/abs/1712.04946}
  {arXiv:1712.04946 [hep-th]} \BibitemShut {NoStop}%
\bibitem [{\citenamefont {Arkani-Hamed}\ \emph {et~al.}(2016)\citenamefont
  {Arkani-Hamed}, \citenamefont {Cohen}, \citenamefont {D'Agnolo},
  \citenamefont {Hook}, \citenamefont {Kim},\ and\ \citenamefont
  {Pinner}}]{Arkani-Hamed:2016rle}%
  \BibitemOpen
  \bibfield  {author} {\bibinfo {author} {\bibfnamefont {N.}~\bibnamefont
  {Arkani-Hamed}}, \bibinfo {author} {\bibfnamefont {T.}~\bibnamefont {Cohen}},
  \bibinfo {author} {\bibfnamefont {R.~T.}\ \bibnamefont {D'Agnolo}}, \bibinfo
  {author} {\bibfnamefont {A.}~\bibnamefont {Hook}}, \bibinfo {author}
  {\bibfnamefont {H.~D.}\ \bibnamefont {Kim}}, \ and\ \bibinfo {author}
  {\bibfnamefont {D.}~\bibnamefont {Pinner}},\ }\href {\doibase
  10.1103/PhysRevLett.117.251801} {\bibfield  {journal} {\bibinfo  {journal}
  {Phys. Rev. Lett.}\ }\textbf {\bibinfo {volume} {117}},\ \bibinfo {pages}
  {251801} (\bibinfo {year} {2016})},\ \Eprint
  {http://arxiv.org/abs/1607.06821} {arXiv:1607.06821 [hep-ph]} \BibitemShut
  {NoStop}%
\bibitem [{\citenamefont {Klevers}\ \emph {et~al.}(2015)\citenamefont
  {Klevers}, \citenamefont {Mayorga~Pe{\~n}a}, \citenamefont {Oehlmann},
  \citenamefont {Piragua},\ and\ \citenamefont {Reuter}}]{Klevers:2014bqa}%
  \BibitemOpen
  \bibfield  {author} {\bibinfo {author} {\bibfnamefont {D.}~\bibnamefont
  {Klevers}}, \bibinfo {author} {\bibfnamefont {D.~K.}\ \bibnamefont
  {Mayorga~Pe{\~n}a}}, \bibinfo {author} {\bibfnamefont {P.-K.}\ \bibnamefont
  {Oehlmann}}, \bibinfo {author} {\bibfnamefont {H.}~\bibnamefont {Piragua}}, \
  and\ \bibinfo {author} {\bibfnamefont {J.}~\bibnamefont {Reuter}},\ }\href
  {\doibase 10.1007/JHEP01(2015)142} {\bibfield  {journal} {\bibinfo  {journal}
  {JHEP}\ }\textbf {\bibinfo {volume} {01}},\ \bibinfo {pages} {142} (\bibinfo
  {year} {2015})},\ \Eprint {http://arxiv.org/abs/1408.4808} {arXiv:1408.4808
  [hep-th]} \BibitemShut {NoStop}%
\bibitem [{\citenamefont {Morrison}\ and\ \citenamefont
  {Vafa}(1996)}]{Morrison:1996pp}%
  \BibitemOpen
  \bibfield  {author} {\bibinfo {author} {\bibfnamefont {D.~R.}\ \bibnamefont
  {Morrison}}\ and\ \bibinfo {author} {\bibfnamefont {C.}~\bibnamefont
  {Vafa}},\ }\href {\doibase 10.1016/0550-3213(96)00369-0} {\bibfield
  {journal} {\bibinfo  {journal} {Nucl. Phys.}\ }\textbf {\bibinfo {volume}
  {B476}},\ \bibinfo {pages} {437} (\bibinfo {year} {1996})},\ \Eprint
  {http://arxiv.org/abs/hep-th/9603161} {arXiv:hep-th/9603161 [hep-th]}
  \BibitemShut {NoStop}%
\bibitem [{\citenamefont {Park}(2012)}]{Park:2011ji}%
  \BibitemOpen
  \bibfield  {author} {\bibinfo {author} {\bibfnamefont {D.~S.}\ \bibnamefont
  {Park}},\ }\href {\doibase 10.1007/JHEP01(2012)093} {\bibfield  {journal}
  {\bibinfo  {journal} {JHEP}\ }\textbf {\bibinfo {volume} {01}},\ \bibinfo
  {pages} {093} (\bibinfo {year} {2012})},\ \Eprint
  {http://arxiv.org/abs/1111.2351} {arXiv:1111.2351 [hep-th]} \BibitemShut
  {NoStop}%
\bibitem [{\citenamefont {Morrison}\ and\ \citenamefont
  {Park}(2012)}]{Morrison:2012ei}%
  \BibitemOpen
  \bibfield  {author} {\bibinfo {author} {\bibfnamefont {D.~R.}\ \bibnamefont
  {Morrison}}\ and\ \bibinfo {author} {\bibfnamefont {D.~S.}\ \bibnamefont
  {Park}},\ }\href {\doibase 10.1007/JHEP10(2012)128} {\bibfield  {journal}
  {\bibinfo  {journal} {JHEP}\ }\textbf {\bibinfo {volume} {10}},\ \bibinfo
  {pages} {128} (\bibinfo {year} {2012})},\ \Eprint
  {http://arxiv.org/abs/1208.2695} {arXiv:1208.2695 [hep-th]} \BibitemShut
  {NoStop}%
\bibitem [{\citenamefont {Cox}\ and\ \citenamefont {Zucker}(1979)}]{Cox1979}%
  \BibitemOpen
  \bibfield  {author} {\bibinfo {author} {\bibfnamefont {D.~A.}\ \bibnamefont
  {Cox}}\ and\ \bibinfo {author} {\bibfnamefont {S.}~\bibnamefont {Zucker}},\
  }\href {\doibase 10.1007/BF01403189} {\bibfield  {journal} {\bibinfo
  {journal} {Inventiones mathematicae}\ }\textbf {\bibinfo {volume} {53}},\
  \bibinfo {pages} {1} (\bibinfo {year} {1979})}\BibitemShut {NoStop}%
\bibitem [{\citenamefont {Lee}\ \emph {et~al.}(2018)\citenamefont {Lee},
  \citenamefont {Regalado},\ and\ \citenamefont {Weigand}}]{Lee:2018ihr}%
  \BibitemOpen
  \bibfield  {author} {\bibinfo {author} {\bibfnamefont {S.-J.}\ \bibnamefont
  {Lee}}, \bibinfo {author} {\bibfnamefont {D.}~\bibnamefont {Regalado}}, \
  and\ \bibinfo {author} {\bibfnamefont {T.}~\bibnamefont {Weigand}},\ }\href
  {\doibase 10.1007/JHEP11(2018)147} {\bibfield  {journal} {\bibinfo  {journal}
  {JHEP}\ }\textbf {\bibinfo {volume} {11}},\ \bibinfo {pages} {147} (\bibinfo
  {year} {2018})},\ \Eprint {http://arxiv.org/abs/1803.07998} {arXiv:1803.07998
  [hep-th]} \BibitemShut {NoStop}%
\bibitem [{\citenamefont {Lee}\ and\ \citenamefont
  {Weigand}(2019)}]{Lee:2019skh}%
  \BibitemOpen
  \bibfield  {author} {\bibinfo {author} {\bibfnamefont {S.-J.}\ \bibnamefont
  {Lee}}\ and\ \bibinfo {author} {\bibfnamefont {T.}~\bibnamefont {Weigand}},\
  }\href {\doibase 10.1103/PhysRevD.100.026015} {\bibfield  {journal} {\bibinfo
   {journal} {Phys. Rev.}\ }\textbf {\bibinfo {volume} {D100}},\ \bibinfo
  {pages} {026015} (\bibinfo {year} {2019})},\ \Eprint
  {http://arxiv.org/abs/1905.13213} {arXiv:1905.13213 [hep-th]} \BibitemShut
  {NoStop}%
\bibitem [{\citenamefont {Cveti{\v c}}\ and\ \citenamefont
  {Lin}(2018{\natexlab{b}})}]{Cvetic:2017epq}%
  \BibitemOpen
  \bibfield  {author} {\bibinfo {author} {\bibfnamefont {M.}~\bibnamefont
  {Cveti{\v c}}}\ and\ \bibinfo {author} {\bibfnamefont {L.}~\bibnamefont
  {Lin}},\ }\href {\doibase 10.1007/JHEP01(2018)157} {\bibfield  {journal}
  {\bibinfo  {journal} {JHEP}\ }\textbf {\bibinfo {volume} {01}},\ \bibinfo
  {pages} {157} (\bibinfo {year} {2018}{\natexlab{b}})},\ \Eprint
  {http://arxiv.org/abs/1706.08521} {arXiv:1706.08521 [hep-th]} \BibitemShut
  {NoStop}%
\bibitem [{\citenamefont {Denef}\ and\ \citenamefont
  {Douglas}(2007)}]{Denef:2006ad}%
  \BibitemOpen
  \bibfield  {author} {\bibinfo {author} {\bibfnamefont {F.}~\bibnamefont
  {Denef}}\ and\ \bibinfo {author} {\bibfnamefont {M.~R.}\ \bibnamefont
  {Douglas}},\ }\href {\doibase 10.1016/j.aop.2006.07.013} {\bibfield
  {journal} {\bibinfo  {journal} {Annals Phys.}\ }\textbf {\bibinfo {volume}
  {322}},\ \bibinfo {pages} {1096} (\bibinfo {year} {2007})},\ \Eprint
  {http://arxiv.org/abs/hep-th/0602072} {arXiv:hep-th/0602072 [hep-th]}
  \BibitemShut {NoStop}%
\bibitem [{\citenamefont {Cveti{\v c}}\ \emph {et~al.}(2011)\citenamefont
  {Cveti{\v c}}, \citenamefont {Garcia-Etxebarria},\ and\ \citenamefont
  {Halverson}}]{Cvetic:2010ky}%
  \BibitemOpen
  \bibfield  {author} {\bibinfo {author} {\bibfnamefont {M.}~\bibnamefont
  {Cveti{\v c}}}, \bibinfo {author} {\bibfnamefont {I.}~\bibnamefont
  {Garcia-Etxebarria}}, \ and\ \bibinfo {author} {\bibfnamefont
  {J.}~\bibnamefont {Halverson}},\ }\href {\doibase 10.1002/prop.201000093}
  {\bibfield  {journal} {\bibinfo  {journal} {Fortsch. Phys.}\ }\textbf
  {\bibinfo {volume} {59}},\ \bibinfo {pages} {243} (\bibinfo {year} {2011})},\
  \Eprint {http://arxiv.org/abs/1009.5386} {arXiv:1009.5386 [hep-th]}
  \BibitemShut {NoStop}%
\bibitem [{\citenamefont {Halverson}\ and\ \citenamefont
  {Ruehle}(2019)}]{Halverson:2018cio}%
  \BibitemOpen
  \bibfield  {author} {\bibinfo {author} {\bibfnamefont {J.}~\bibnamefont
  {Halverson}}\ and\ \bibinfo {author} {\bibfnamefont {F.}~\bibnamefont
  {Ruehle}},\ }\href {\doibase 10.1103/PhysRevD.99.046015} {\bibfield
  {journal} {\bibinfo  {journal} {Phys. Rev.}\ }\textbf {\bibinfo {volume}
  {D99}},\ \bibinfo {pages} {046015} (\bibinfo {year} {2019})},\ \Eprint
  {http://arxiv.org/abs/1809.08279} {arXiv:1809.08279 [hep-th]} \BibitemShut
  {NoStop}%
\bibitem [{\citenamefont {Halverson}\ \emph {et~al.}(2020)\citenamefont
  {Halverson}, \citenamefont {Plesser}, \citenamefont {Ruehle},\ and\
  \citenamefont {Tian}}]{Halverson:2019vmd}%
  \BibitemOpen
  \bibfield  {author} {\bibinfo {author} {\bibfnamefont {J.}~\bibnamefont
  {Halverson}}, \bibinfo {author} {\bibfnamefont {M.}~\bibnamefont {Plesser}},
  \bibinfo {author} {\bibfnamefont {F.}~\bibnamefont {Ruehle}}, \ and\ \bibinfo
  {author} {\bibfnamefont {J.}~\bibnamefont {Tian}},\ }\href {\doibase
  10.1103/PhysRevD.101.046010} {\bibfield  {journal} {\bibinfo  {journal}
  {Phys. Rev.}\ }\textbf {\bibinfo {volume} {D101}},\ \bibinfo {pages} {046010}
  (\bibinfo {year} {2020})},\ \Eprint {http://arxiv.org/abs/1911.07835}
  {arXiv:1911.07835 [hep-th]} \BibitemShut {NoStop}%
\bibitem [{\citenamefont {Ooguri}\ and\ \citenamefont
  {Vafa}(2007)}]{Ooguri:2006in}%
  \BibitemOpen
  \bibfield  {author} {\bibinfo {author} {\bibfnamefont {H.}~\bibnamefont
  {Ooguri}}\ and\ \bibinfo {author} {\bibfnamefont {C.}~\bibnamefont {Vafa}},\
  }\href {\doibase 10.1016/j.nuclphysb.2006.10.033} {\bibfield  {journal}
  {\bibinfo  {journal} {Nucl. Phys.}\ }\textbf {\bibinfo {volume} {B766}},\
  \bibinfo {pages} {21} (\bibinfo {year} {2007})},\ \Eprint
  {http://arxiv.org/abs/hep-th/0605264} {arXiv:hep-th/0605264 [hep-th]}
  \BibitemShut {NoStop}%
\bibitem [{\citenamefont {Grimm}\ \emph {et~al.}(2018)\citenamefont {Grimm},
  \citenamefont {Palti},\ and\ \citenamefont {Valenzuela}}]{Grimm:2018ohb}%
  \BibitemOpen
  \bibfield  {author} {\bibinfo {author} {\bibfnamefont {T.~W.}\ \bibnamefont
  {Grimm}}, \bibinfo {author} {\bibfnamefont {E.}~\bibnamefont {Palti}}, \ and\
  \bibinfo {author} {\bibfnamefont {I.}~\bibnamefont {Valenzuela}},\ }\href
  {\doibase 10.1007/JHEP08(2018)143} {\bibfield  {journal} {\bibinfo  {journal}
  {JHEP}\ }\textbf {\bibinfo {volume} {08}},\ \bibinfo {pages} {143} (\bibinfo
  {year} {2018})},\ \Eprint {http://arxiv.org/abs/1802.08264} {arXiv:1802.08264
  [hep-th]} \BibitemShut {NoStop}%
\bibitem [{\citenamefont {Lee}\ \emph {et~al.}(2019)\citenamefont {Lee},
  \citenamefont {Lerche},\ and\ \citenamefont {Weigand}}]{Lee:2019wij}%
  \BibitemOpen
  \bibfield  {author} {\bibinfo {author} {\bibfnamefont {S.-J.}\ \bibnamefont
  {Lee}}, \bibinfo {author} {\bibfnamefont {W.}~\bibnamefont {Lerche}}, \ and\
  \bibinfo {author} {\bibfnamefont {T.}~\bibnamefont {Weigand}},\ }\href@noop
  {} {\  (\bibinfo {year} {2019})},\ \Eprint {http://arxiv.org/abs/1910.01135}
  {arXiv:1910.01135 [hep-th]} \BibitemShut {NoStop}%
\end{thebibliography}%

\end{document}